\documentclass[prx, twocolumn,superscriptaddress]{revtex4-1}

\usepackage{amsmath,amssymb, amsfonts}
\usepackage{bm}
\usepackage{graphicx}
\usepackage{braket}
\usepackage{ascmac}
\usepackage{color}
\usepackage{multirow}

\newcommand{\pmat}[1]{\begin{pmatrix} #1 \end{pmatrix}}%

\newcommand{\dd}{\mathrm{d}}
\newcommand{\tr}{\mathrm{tr}}

\begin{document}
\title{Nielsen-Ninomiya Theorem with Bulk Topology: \\
Duality in Floquet and non-Hermitian Systems}

\author{Takumi Bessho}
\email{takumi.bessho@yukawa.kyoto-u.ac.jp}
\affiliation{Yukawa Institute for Theoretical Physics, Kyoto University, Kyoto 606-8502, Japan}

\author{Masatoshi Sato}
\email{msato@yukawa.kyoto-u.ac.jp}
\affiliation{Yukawa Institute for Theoretical Physics, Kyoto University, Kyoto 606-8502, Japan}

\date{\today}

\begin{abstract}
The Nielsen-Ninomiya theorem is a fundamental theorem on the realization of chiral fermions in static lattice systems in high-energy and condensed matter physics. Here we extend the theorem in {\sl dynamical systems}, which include the original Nielsen-Ninomiya theorem in the static limit. In contrast to the original theorem, which is a no-go theorem for bulk chiral fermions, the new theorem permits them due to bulk topology intrinsic to dynamical systems. The theorem is based on duality enabling a unified treatment of periodically driven systems and non-Hermitian ones.
We also present the extended theorem for non-chiral gapless fermions protected by symmetry.
Finally, as an application of our theorem and duality, we predict a new type of chiral magnetic effect --- the non-Hermitian chiral magnetic skin effect. 
\end{abstract}

\maketitle

The Nielsen-Ninomiya (NN) theorem is a fundamental constraint in realizing chiral fermions in lattice systems \cite{Nielsen81, Nielsen81ii, Karsten81}. 
It initially was a no-go theorem for the lattice realization of the Standard Model in particle physics, but it also applies to condensed matter physics. For instance, the Nielsen-Ninomiya theorem requires that bulk Weyl points in Weyl semimetals always appear in a pair so that the total chiral charge of Weyl points vanishes \cite{Nielsen83, Murakami07, Wan11}.
The NN theorem severely restricts bulk low energy modes in topological materials \cite{Burkov11, Zyuzin12, Vazifeh13, Xu15, Armitage18, Chiu14, Kobayashi14, Fang15}.

However, recent studies have revealed that the NN theorem does not hold when considering topological states in dynamical systems \cite{Jiang11, Kitagawa11, Kitagawa12, Rudner13, Asboth13, Nathan15, Carpentier15, Roy15, Zhou16, Morimoto17, Nakagawa20,Hatano, Dembowski01, Rudner09, SHEK12, Hu, Esaki11, Schomerus13, Zeuner, Lee16, Leykam17, Xiao17, Chernodub17, Shen18, ZL2019, Yao18, Yao18t, Kunst18, Kawabata18, Xu18, Yoshida18, Zyuzin18, Longhi19, Okugawa19, KBS, OKuma19, Song19, Ge19, Yoshida19, Song19-2, Imura19, Kimura19, Rui19, Matsushita19, Moors19, Yang19, Jin19, Zhou20, Ohashi20, Longhi20, Yang20, Wojcik20, Yang19-2, Kawabata20, Terrier20, Chernodub20, Borgnia20}: Periodically driven systems may support unpaired chiral fermions both in one-\cite{Kitagawa10, Titum16, Budich17, Privitera18, Wauters19} and three-dimensions\cite{Sun18, Higashikawa18}. Furthermore, systems with non-Hermitian Hamiltonians also retain unpaired chiral fermions after the long-time dynamics \cite{Lee19}.
These examples have suggested a reformulation of the NN theorem in dynamical systems.

In this Letter, we extend the NN theorem in dynamical systems. As a particular case of the static limit, the extended theorem includes the original one. A key of our extension is a duality between periodically driven systems and non-Hermitian ones. A one-cycle time evolution operator $U_{\rm F}$ generally describes a periodically driven system. By identifying $iU_{\rm F}$ as a non-Hermitian Hamiltonian $H$, we treat a periodically driven system and a non-Hermitian one in a unified manner.
Another key is multiple gap structures intrinsic to non-Hermitian systems. The complex energy spectrum of non-Hermitian systems may introduce two different gap structures: point and line gaps \cite{Gong18,KSUS18}. 
A non-Hermitian system can be gapped in the sense of point gap even if it supports gapless fermions in the sense of line gap. Because the point gap enables a novel bulk topological number, this means that bulk chiral (so gapless) fermions in dynamical systems may coexist with non-trivial bulk topology. 
This situation never happens in conventional static systems and makes it possible to reformulate the NN theorem.

The extended NN theorem provides an exact relation between the total chiral charge of chiral fermions and the bulk topological number. This theorem infers that if the bulk topological number is nonzero, so is the total chiral charge, and thus the system realizes unpaired chiral fermions. The extended theorem also applies to systems with symmetry. Symmetry protects non-chiral gapless fermions, giving them a topological charge other than chirality. In this case, the bulk topological number is equal to the total topological charge from our theorem.

As an application of our theorem, we consider a non-Hermitian version of the chiral magnetic effect (CME). The CME is an electric current generation along an applied magnetic field due to unpaired Weyl fermions in three dimensions \cite{Fukushima08}. While the chiral magnetic effect does not occur in static systems because of the NN theorem \cite{Vazifeh13}, the extended theorem allows it in dynamical systems. Periodically driven systems may exhibit the CME \cite{Sun18, Higashikawa18}, and thus our duality relation suggests that so do non-Hermitian systems. We demonstrate that a wave packet in a non-Hermitian Weyl semimetal moves in the direction of an applied magnetic field, manifesting the CME. Furthermore, the extended theorem implies a nonzero spectral winding number of non-Hermitian Weyl semimetals under a magnetic field. This result leads to predicting a new type of CME---the chiral magnetic skin effect.

We assume without loss of generality that the Fermi energy $E_{\rm F}$, {\it i.e.} the reference energy of a gap, is zero unless otherwise mentioned.
One can recover $E_{\rm F}$ by replacing the Hamiltonian $H({\bm k})$ with $H({\bm k})-E_{\rm F}$ if necessary.

{\it 1D chiral fermions in dynamical systems.---}
Let us start with a simple 1D non-Hermitian system hosting a chiral mode. 
The Hamiltonian of the model is 
\begin{align}
H(k)=\sin k +i \cos k,    
\label{eq:1dnon-Hermitian}
\end{align}
where $k$ is the crystal momentum and $H(k)$ is periodic in $k$ \cite{Lee19}.  
The energy $E(k)$ of the system is $H(k)$ itself, and the group velocity $v(k)$ 
is
$
v(k)={\rm Re}(\partial E(k)/\partial k).   
$
At the Fermi energy ${\rm Re}E(k)=0$, there are two gapless modes with $k=0,\pi$:
A right-moving mode $(v(k)>0)$ with $k=0$ and a left-moving mode $(v(k)<0)$ with $k=\pi$. While the right-moving mode has a positive ${\rm Im}E(k)$, the left-moving mode has a negative one; thus, the left-moving mode decays, and only the right-moving mode survives after the long-time dynamics. Therefore, the system realizes a chiral fermion, {\it i.e.} a right-moving chiral mode. 

Another simple 1D model with a chiral mode is a periodically driven system evolved by the one-component unitary operator \cite{Sun18},
\begin{align}
U_{\rm F}(k)=e^{-ik}.    
\label{eq:1dFloquet}
\end{align}
The Floqet Hamiltonian $H_{\rm F}(k)$ defined by
$e^{-iH_{\rm F}(k)\tau}=U_{\rm F}(k)$ with 
a driving period $\tau$ 
describes the stroboscopic time-evolution of the system, 
$|t+\tau\rangle=U_{\rm F}(t)|t\rangle=e^{-iH_{\rm F}(k)\tau}|t\rangle$.
The eigenvalue of $H_{\rm F}(k)$, called the quasi-energy, is $\epsilon_{\rm F}(k)=k/\tau$ up to an integer multiple of $2\pi/\tau$. Because the group velocity $v_{\rm F}(k)=\partial \epsilon_{\rm F}(k)/\partial k$ is positive, the system has a right-moving chiral mode.

These chiral modes have a common topological origin. 
The equation 
\begin{align}
H(k)=iU_{\rm F}(k),   
\label{eq:duality1d}
\end{align}
relates the above models,
then the 1D (spectral) winding number 
\begin{align}
w_1=
-\int_0^{2\pi}\frac{dk}{2\pi i}
{\rm tr}[H^{-1}(k)\partial_k H(k)].
\label{eq:w1}
\end{align}
gives $w_1=1$ for both models.
(The trace is trivial in the above models.)
For the non-Hermitian model in Eq.(\ref{eq:1dnon-Hermitian}), the non-zero spectral winding number results in so-called the non-Hermitian skin effect \cite{Yao18}: For $w_1=1$, all bulk states localize to the right end \cite{ZYF19,OKSS20}. This effect suggests a right-moving chiral mode because a uni-directed movement of the mode forces
all bulk states to move to the right end. 
For the periodically driven model in Eq.(\ref{eq:1dFloquet}), on the other hand, the non-zero spectral winding number implies a non-zero average of the group velocity,
\begin{align}
w_1=-\int_0^{2\pi}\frac{dk}{2\pi i}
\partial_k \ln {\rm det}H(k)=
\int_0^{2\pi}\frac{dk}{2\pi}v_{\rm F}(k)\tau,  
\end{align}
which also indicates a right-moving chiral mode. 

\begin{figure}[htbp]
\centering
\includegraphics[width=86mm]{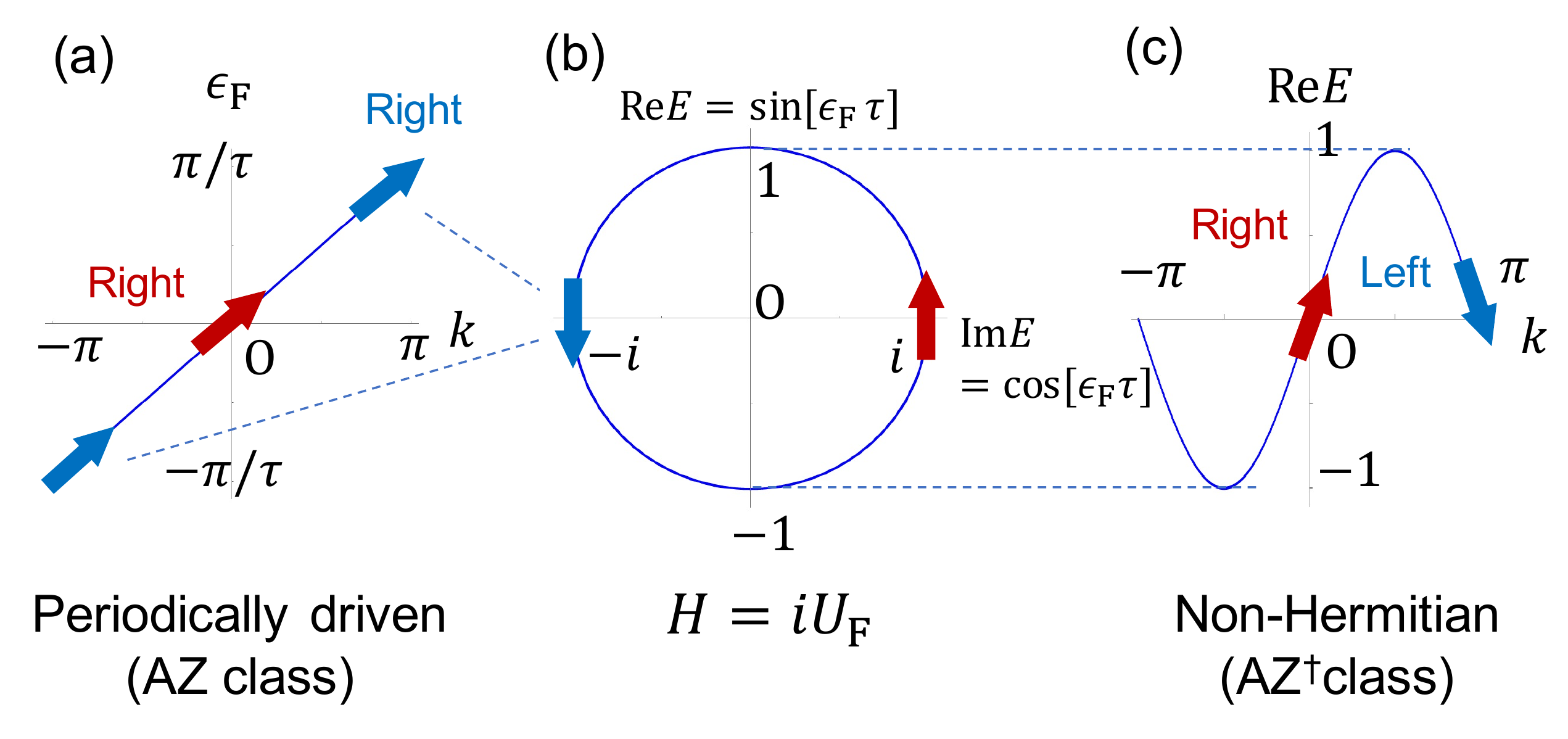} 
\caption{Duality between a periodically driven system and a non-Hermitian one. We illustrate the 1D case here. $w_1$ is the winding number of the spectral in the complex energy plane in (b). Theorem 1' is evident in the relation between (a) and (b).
The duality holds in any dimensions.}
	\label{fig: NHWeyl}
\end{figure}

The above examples suggest a general relation between the spectral winding number and the chirality ${\rm sgn}\,v_{\rm F}(k)$ of gapless modes. For 1D non-Hermitian systems, the exact link is as follows \footnote{See Sec.\ref{sec:S2} in Supplemental Material, which includes Refs.  \cite{Sato09,QTZ10,STYY11}.}:

{\bf Theorem 1}:
Let $H(k)$ be a 1D non-Hermitian Hamiltonian and $E_p(k)$ be the complex eigen-energy of band $p$.  
Then, we have
\begin{align}
w_1=\sum_{{\rm Im}E_p(k_{p\alpha})>0}{\rm sgn}\,v_{p\alpha}=
-\sum_{{\rm Im}E_p(k_{p\alpha})<0}{\rm sgn}\,v_{p\alpha},
\label{eq:1dENN}
\end{align}
where $k_{p\alpha}$ is the $\alpha$-th Fermi point of band $p$ defined by ${\rm Re}E_{p}(k_{p\alpha})=0$, and $v_{p\alpha}={\rm Re}(\partial E_p(k)/\partial k)_{k=k_{p\alpha}}$ is the group velocity at $k_{p\alpha}$.
The summation in Eq.(\ref{eq:1dENN}) is over all $p$ and $\alpha$.

For a Hermitian Hamiltonian $H(k)$, the above theorem reproduces the NN theorem. 
The spectral winding number $w_1$ is zero for any Hermitian Hamiltonian, and by adding a small imaginary term $i\eta$ to $H(k)$, all the Fermi points can have a positive imaginary part of the energy. 
Thus, from Eq. (\ref{eq:1dENN}), we have $\sum_{k_{p\alpha}} {\rm sgn}\, v_{p\alpha}=0$, which is the NN theorem in one-dimension \cite{Nielsen81}. 

Using the relation in Eq.(\ref{eq:duality1d}), we can also derive a counterpart theorem for 1D periodically driven systems:
Equation (\ref{eq:duality1d}) maps the quasi-energy $\epsilon_p(k)$ of $U_{\rm F}(k)$ to 
the complex energy $E_p(k)$ of $H(k)$, $E_p(k)=\sin [\epsilon_p(k)\tau]+i \cos [\epsilon_p(k)\tau]$.
Thus, a Fermi point defined by $\epsilon_p(k)=0$ $(\pi/\tau)$ gives a Fermi point of $E_p(k)$ with a positive (negative) ${\rm Im}E_p(k)$. Comparing the group velocities at the Fermi points, we obtain the theorem: 

{\bf Theorem 1'}: Let $H_{\rm F}(k)$ be a 1D Floquet Hamiltonian and $\epsilon_p(k)$ be the quasi-energy of band $p$. Then, gapless modes of the quasi energy obey
\begin{align}
w_1=\sum_{\epsilon_p(k_{p\alpha})=\mu}{\rm sgn}\,v_{p\alpha}, 
\label{eq:1dNNFloquet}
\end{align}
where $k_{p\alpha}$ is the Fermi point of band $p$ defined by $\epsilon(k_{p\alpha})=\mu$, and $v_{p\alpha}=(\partial \epsilon_p(k)/\partial k)_{k=k_{p\alpha}}$ is the group velocity at $k_{p\alpha}$ \footnote{This result is known as the Brouwer degree in mathematics, and was shown in Ref.\cite{Sun18} in a different manner.}.

Here we have shifted the origin of the quasi-energy by $U_{\rm F}\rightarrow e^{i\mu\tau}U_{\rm F}$ and omitted the term corresponding to the last term in Eq.(\ref{eq:1dENN}) since it is just a particular case of Eq.(\ref{eq:1dNNFloquet}).


{\it Non-Hermitian Weyl semimetals---}
Weyl fermions are 3D massless (or gapless) fermions with a definite chirality. They are realized as band crossing points (Weyl points) and behave like magnetic monopoles in the momentum space,  of which the magnetic charge provides the chirality charge.
They have finite lifetimes --the imaginary part of the energies-- in the presence of non-Hermiticity.  For Weyl fermions, we have the following theorem \footnote{See Sec.\ref{sec:S2} in Supplemental Material.}

{\bf Theorem 2}:
Let $H({\bm k})$ be a 3D non-Hermitian Hamiltonian and $E_p({\bm k})$ be the complex eigen-energy of band $p$.  
Then, Weyl fermions in the complex energy spectrum obey
\begin{align}
w_3=\sum_{{\rm Im}E_p(S_{p\alpha})>0}{\rm Ch}_{p\alpha}=
-\sum_{{\rm Im}E_p(S_{p\alpha})<0}{\rm Ch}_{p\alpha}.
\end{align}
Here $w_3$ is the 3D winding number, 
\begin{align}
w_3=-\frac{1}{24\pi^2}\int_{\rm BZ}{\rm tr}[H^{-1}dH]^3,  
\label{eq:w3}
\end{align}
$S_{p\alpha}$ is the $\alpha$-th Fermi surface of band $p$ defined by $S_{p\alpha}=\{{\bm k}\in {\rm BZ}|{\rm Re}E_p({\bm k})=0\}$, and ${\rm Ch}_{p\alpha}$ is the Chern number on the Fermi surface $S_{p\alpha}$,
\begin{align}
{\rm Ch}_{p\alpha}=\frac{1}{2\pi i}\int_{S_{p\alpha}}  (\nabla \times \bm{A}(\bm{k}) )\cdot \text{d}{\bm S},
\end{align}
where ${\bm A}({\bm k})=\langle\!\langle \psi_p({\bm k})|\nabla \psi_p({\bm k})\rangle$ with $H({\bm k})|\psi_p({\bm k})\rangle=E_p({\bm k})|\psi_p({\bm k})\rangle$, $H^{\dagger}({\bm k})|\psi_p({\bm k})\rangle\!\rangle=E^*_p({\bm k})|\psi_p({\bm k})\rangle\!\rangle$, and the orientation of $S_{p\alpha}$ is along the direction of the Fermi velocity ${\rm Re}(\partial E_p({\bm k})/\partial {\bm k})_{{\bm k}\in S_{p\alpha}}$.
${\rm Ch}_{p\alpha}$ counts the total chirality of Weyl points inside $S_{p\alpha}$.

Theorem 2 reproduces the NN theorem again when $H({\bm k})$ is Hermitian: By adding a tiny positive imaginary term to $H(k)$, we have $\sum_{p\alpha}{\rm Ch}_{p\alpha}=0$, 
which is one of the variants of the NN theorem in three dimensions \footnote{See Sec.\ref{sec:S3} in Supplemental Material.}.
Indeed, this equation forbids an unpaired Weyl point in Hermitian systems:
If an unpaired Weyl point were to exist, we would have a Fermi surface surrounding it by choosing the Fermi energy near the Weyl point.  
This configuration would give a nonzero $\sum_{p\alpha}{\rm Ch}_{p\alpha}$, which contradicts $\sum_{p\alpha}{\rm Ch}_{p\alpha}=0$.

When $w_3$ is nonzero, Theorem 2 predicts chiral fermions. 
For instance, consider the following model,
\begin{align}\label{eq:BesWeyl}
H({\bm k})&=\left(d_0+{\bm d}({\bm k})\cdot{\bm \sigma}\right) \tau_{1} +m({\bm k})\tau_3+i\gamma(\tau_3
-\tau_0),
\end{align}
with
$d_i({\bm k})=\sin k_i$, $m({\bm k})=m_0+\sum_{i=1}^3\cos k_i$. This model has a point gap at $E_{\rm F}=-i\gamma$ and hosts Weyl points in the complex energy plane as shown in Fig.~\ref{fig:BesWeylEnergy}(a),  satisfying Theorem 2 \footnote{See Sec.\ref{sec:S4} in Supplemental Material.}.

{\it Duality.---}
The relation (\ref{eq:duality1d}), which
enables a unified treatment of a periodically driven system and a non-Hermitian one, is not accidental.
This duality relation holds in arbitrary dimensions.
Evidently, one can immediately identify any one-cycle time evolution operator $U_{\rm F}({\bm k})$ with a non-Hermitian Hamiltonian $H({\bm k})$ by
\begin{align}
H({\bm k})=i U_{\rm F}({\bm k}). 
\label{eq:duality}
\end{align}
However, the opposite is also true for a class of non-Hermitian systems.
We say that a non-Hermitian Hamiltonian $H({\bm k})$ 
has a point gap if ${\rm det}H({\bm k})\neq 0$.
Then, one can regard any point-gapped Hamiltonian as a one-cycle time evolution operator
because a point gapped $H({\bm k})$ can smoothly deform into a unitary matrix without closing the point gap \cite{Gong18,KSUS18}.

The duality relation (\ref{eq:duality}) brings out common properties of periodically driven systems and non-Hermitian ones:
In terms of the Floquet Hamiltonian $H_{\rm F}({\bm k})=(i/\tau)\ln U_{\rm F}({\bm k})$, 
the above relation reads
$
H({\bm k})=\sin [H_{\rm F}({\bm k})\tau]
+i\cos [H_{\rm F}({\bm k})\tau].
$
Thus, eigenstates of $H({\bm k})$ are identical to those of $H_{\rm F}({\bm k})$.
Also, a gapless state in $H_{\rm F}({\bm k})\sim {\bm k}\cdot{\bm \Gamma}$ results in
a gapless state in $H({\bm k})$, and vice versa. (${\bm \Gamma}$ are some matrices.)
Furthermore, these systems share a topological number; the topological number is given by that of the Hermitian Hamiltonian \cite{Gong18, Higashikawa18, KSUS18},
\begin{align}
{\cal H}({\bm k})=
\begin{pmatrix}
0 & H({\bm k})\\
H^{\dagger}({\bm k}) & 0
\end{pmatrix}.
\label{eq:doubleH}
\end{align}
From Eq.(\ref{eq:duality}), ${\cal H}({\bm k})$ 
satisfies ${\cal H}^2({\bm k})=1$ and thus has eigenvalues $\pm 1$.
Therefore, ${\cal H}({\bm k})$ defines an insulator,  giving a well-defined topological number.

Note that the above identification links a periodically driven system and a non-Hermitian one in different symmetry classes.  
To see this, consider time-reversal, particle-hole, and chiral symmetries for the Floquet Hamiltonian $H_{\rm F}({\bm k})$, given by
$T H_{\rm F}({\bm k})T^{-1}=H_{\rm F}(-{\bm k})$, $C H_{\rm F}({\bm k}) C^{-1}=-H_{\rm F}(-{\bm k})$, and $\Gamma H_{\rm F}({\bm k})\Gamma^{-1}=-H_{\rm F}({\bm k})$, respectively. Here $T$ and $C$ are antiunitary operators with $T^2=\pm 1$, $C^2=\pm 1$, and $\Gamma$ is a unitary operator with $\Gamma^2=1$.
The presence or absence of these symmetries define  Altland-Zirnbauer (AZ) ten symmetry classes \cite{AZ97}.
The relation (\ref{eq:duality}) maps these symmetries as follows:
$T H^{\dagger}(\bm{k}) T^{-1}=H(-\bm{k})$, $C H(\bm{k}) C^{-1}=-H(-\bm{k})$, and $\Gamma H^{\dagger}(\bm{k}) \Gamma^{-1}=-H(\bm{k})$.
The latter symmetries define another ten symmetry classes, called AZ$^{\dagger}$ classes \cite{KSUS18},
which are intrinsic to non-Hermitian systems.

{\it Extended NN theorem. ---}
Symmetry protects gapless fermions other than chiral fermions.
We now present the extended NN theorem, including such non-chiral (Dirac) fermions.

First, consider non-Hermitian systems.
Depending on symmetry classes, two different situations may happen: (i) 
gapless fermions in classes A, AI$^{\dagger}$, AII$^{\dagger}$ appear as band crossing points at general positions in the complex energy plane, and (ii) those in other AZ$^{\dagger}$ classes appear on the ${\rm Re}E= 0$ axis. 
To define the topological charge of gapless fermions,
we use the Fermi surface at ${\rm Re}E=0$ in the former case, 
and a small sphere encircling a gapless fermion in the latter \footnote{See Sec.\ref{sec:S0} in Supplemental Material}.
We have the following theorem:

{\bf Theorem 3}: 
Let $H({\bm k})$ be a point-gapped non-Hermitian Hamiltonian in an  AZ$^{\dagger}$ class.
Then, bulk gapless fermions of $H({\bm k})$ obeys
\begin{align}\label{eq:Thm1}
n=\sum_{\text{Im}E_\alpha>0}\nu_{\alpha}=-\sum_{\text{Im}E_\alpha<0}\nu_{\alpha},
\end{align}
As we mentioned above, the point gap topological number $n$ is given by the conventional topological number of the topological insulator described by the Hermitian Hamiltonian in Eq.(\ref{eq:doubleH}). 
The explicit form of n is summarized in Ref. \cite{KSUS18}.
In case (i) in the above, $\alpha$ labels the Fermi surfaces at ${\rm Re}E=0$,  $\nu_{\alpha}$ is the topological charge of gapless fermions inside the $\alpha$-th Fermi surface, and $E_\alpha$ is the complex energy of the Fermi surface.
In case (ii), $\alpha$ labels gapless fermions, $\nu_{\alpha}$ is the topological charge of the $\alpha$-th gapless fermion defined on the small sphere, and $E_\alpha$ is the complex energy of the gapless fermion
\footnote{For the precise definition of the topological charge $\nu_{\alpha}$ and the proof of Theorem 3', see Sec.\ref{sec:S2} in Supplemental Material.}.


Using the duality relation (\ref{eq:duality}), we also have an accompanying theorem for gapless fermions in periodically driven systems.
We find that (i') 
gapless fermions in classes A, AI, AII appear as band crossing points with arbitrary energies in the quasi-energy spectra, and (ii') those in other AZ classes appear with $\epsilon=0$ or $\pi/\tau$. Then, the accompanying theorem is as follows.

{\bf Theorem 3'}:
For gapless fermions in a periodically driven system in an AZ class, we have
\begin{align}
  \label{eq:Thm2-1}
n&= \sum_{\epsilon_\alpha=\mu}\nu^\mu_\alpha, \quad &\mbox{in case (i')},
\\
\label{eq:Thm2-2}
n= \sum_{\epsilon_\alpha=0}\nu_{\alpha}^0&=-(-1)^d\sum_{\epsilon_\alpha=\pi/\tau}\nu_{\alpha}^\pi, \quad &\mbox{in case (ii')}.
\end{align}
Here $n$ is the topological number of $iU_{\rm F}({\bm k})$ given by ${\cal H}({\bm k})$ in Eq.(\ref{eq:doubleH}), and $d$ is the dimension of the system. In case (i'), $\alpha$ labels the Fermi surfaces defined by $\epsilon=\mu$, and $\nu_{\alpha}^\mu$ is the topological charge of gapless fermions inside the $\alpha$-th Fermi surface. In case (ii), $\alpha$ labels gapless fermions at $\epsilon=0,\pi$, and $\nu_{\alpha}^{0,\pi}$ is the topological charge of the gapless fermion with the quasi-energy $\epsilon_\alpha=0,\pi$ \footnote{See Sec.\ref{sec:SF} in Supplemental Material. } . 

Note that Eq.(\ref{eq:Thm2-2}) have a sign depending on $d$ in the second equality: A gapless fermion at $\pi/\tau$, $H_{\rm F}({\bm k})={\bm k}\cdot {\bm \Gamma}+\pi/\tau$, in a periodically driven system corresponds to $H({\bm k})=-{\bm k}\cdot {\bm \Gamma}-i$, in a non-Hermitian system.  
Since these Hamiltonians have an opposite topological charge in odd dimensions, we have the additional sign $(-1)^d$. Equation (\ref{eq:1dNNFloquet}) is the 1D case of Eq.(\ref{eq:Thm2-1}) in class A (no symmetry).
We have also confirmed Eq.(\ref{eq:Thm2-2}) using a 2D periodically driven model with chiral symmetry (class AIII) \footnote{See Sec.\ref{sec:S5} in Supplemental Material.}.

{\it Chiral magnetic effect.---}
Weyl fermions in a periodically driven system may exhibit the CME \cite{Higashikawa18, Sun18}.
 As a counterpart of this effect, we investigate the non-Hermitian CME. Figure~\ref{fig:BesWeylEnergy}(b) shows the energy spectrum of the model in Eq.~(\ref{eq:BesWeyl}) under a magnetic field $B_z$ in the $z$-direction. The magnetic field opens the Landau gap at the Weyl point at $k = (0,0,0)$ in Fig.~\ref{fig:BesWeylEnergy}(a), and a right-moving chiral mode with a positive ${\rm Im}(E-E_{\rm F})$ appears. The chiral mode has a longer lifetime and produces a current along the magnetic field, leading to the CME. We confirm the CME by examining the wave packet dynamics. Figures \ref{fig:BesWeylEnergy}(c) and \ref{fig:BesWeylEnergy}(d) show the wave packet dynamics without and with a magnetic field. 
 While wave packets without a magnetic field tend to move along the spin direction because of the spin-momentum locking of Weyl fermions, we observe different uni-directed motions with a magnetic field consistent with the CME.

\begin{figure}[htbp]
\centering
\includegraphics[width=85mm]{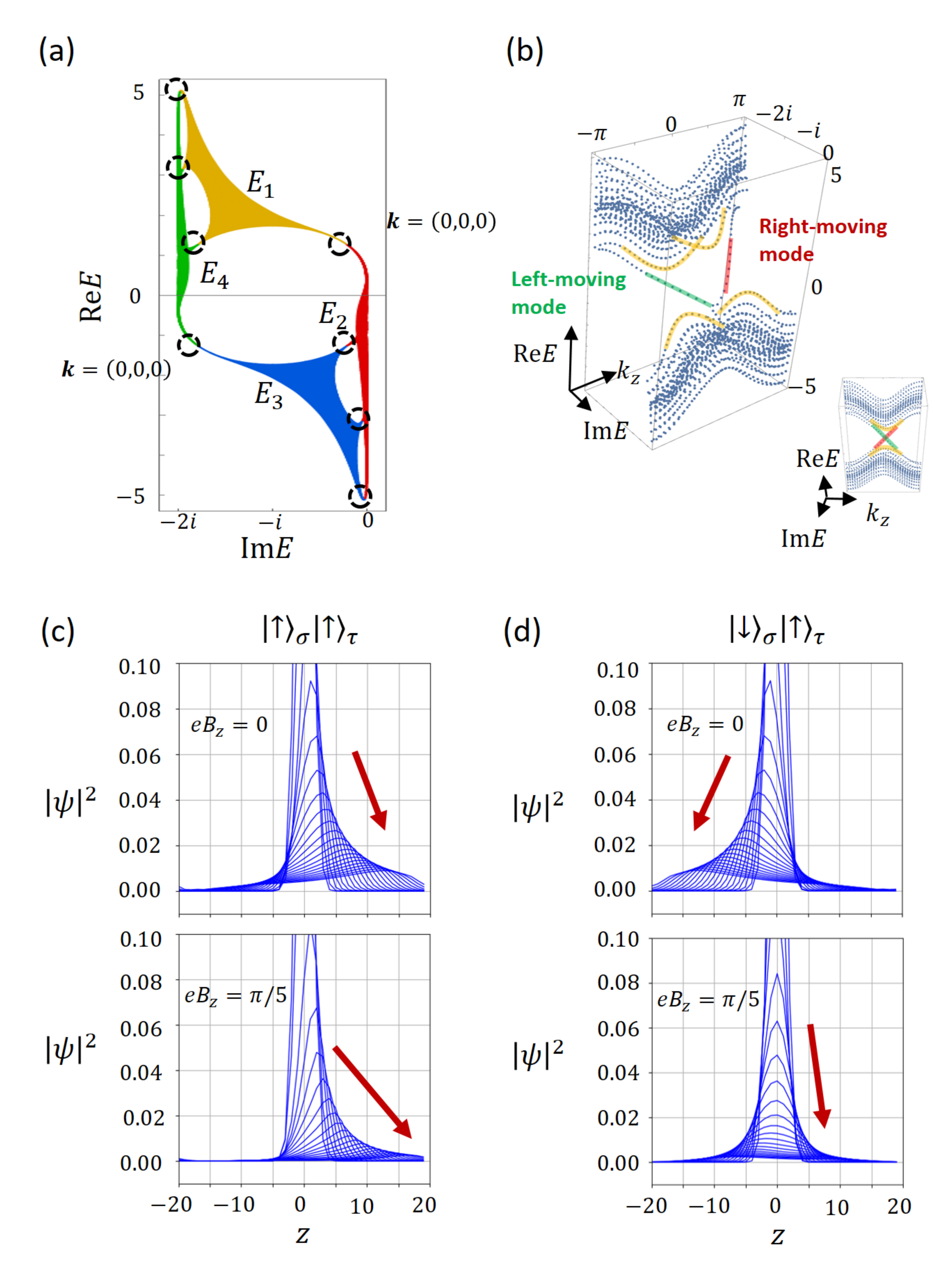} 
\caption{(a, b) Energy spectrum of the non-Hermitian Weyl semimetal in Eq.~(\ref{eq:BesWeyl}) (a) without and (b) with a magnetic field $B_z$ in the $z$ direction. 
(a) Colors distinguish different bands, and dotted circles enclose Weyl points.  
(b) Right (Left) moving mode has positive (negative) ${\rm Im}(E-E_{\rm F})$ with $E_{\rm F}=-i$.
The right and left moving modes originate from Weyl points with ${\bm k}=(0,0,0)$. The inset is the same figure viewed from a different angle.
(c,d) Wave packet dynamics in the non-Hermitian Weyl semimetal of Eq.~(\ref{eq:BesWeyl}) (top) without and (bottom) with $B_z$.
We draw snapshots of the probability densities $|\psi(z)|^2$ at each unit cycle, where the red arrows indicate the time evolution. 
We use the fourth-order Runge-Kutta method. 
The initial wave packets are $\ket{\psi_0}=\psi_0\ket{\sigma_z}_{\sigma}\ket{\tau_z}_{\tau}$, where $\psi_0$ is a 3D Gaussian wave packet with the width $2\bar{\sigma}^2=5$ and $\ket{\sigma_z}_{\sigma}\ket{\tau_z}_{\tau}$ is specified in each figure. 
With $B_z=\pi/5$, all the wave packets tend to move in the $+\hat{z}$ direction.
The parameters in Eq.~(\ref{eq:BesWeyl}) are $d_0=\gamma=\gamma_0=1$ and $m_0=-2$. 
The system size is (b) $L_x=L_y=L_z=30$ and (c,d) $L_x=L_y=L_z=40$ with the periodic boundary conditions.  
}
	\label{fig:BesWeylEnergy}
\end{figure}

Using the extended NN theorem, we can predict a general effect intrinsic to the non-Hermitian CME: From Theorem 2, a system with nonzero $w_3$ hosts Weyl fermions with the total chiral charge of $w_3$.
As in Fig.\ref{fig:BesWeylEnergy} (b),
a magnetic field $B_z$ opens the Landau gap at Weyl fermions, leaving a 1D chiral mode for each Weyl point, with the Landau degeneracy $(eB_z/2\pi)L_xL_y$ \footnote{See Sec.\ref{sec:S4} in Supplemental Material}, where $e$ is the electric charge of the Weyl fermion and $L_{i=x,y}$ is the system length in the $i$-direction. Therefore, the system supports 1D chiral modes with the total chiral charge $w_3(eB_z/2\pi)L_xL_y$.  From Theorem 1, this result implies that the system also hosts the 1D spectral winding number $w_1$ given by
\begin{align}\label{eq:relation}
w_1=\frac{eB_z}{2\pi} L_x L_y w_3.
\end{align}
Here $w_1$ is defined by Eq.(\ref{eq:w1}), where $H(k)$ with $k=k_z$ is
the Hamiltonian under the magnetic field $B_z$, and the trace includes the summation of $k_x$ and $k_y$ in the magnetic Brillouin zone. 
Note that $eB_zL_xL_y/2\pi$ is an integer under the periodic boundary conditions in $x$- and $y$-directions.

The relation (\ref{eq:relation}) gives a profound implication. As mentioned above, a nonzero $w_1$ induces the non-Hermitian skin effect \cite{ZYF19, OKSS20}, where extended bulk modes in the periodic boundary condition become localized boundary modes in the open boundary condition. Therefore, Eq. (\ref{eq:relation}) predicts that the system with a nonzero $w_3$ inevitably shows the skin effect under a magnetic field. This prediction is consistent with the CME because bulk modes stack to a boundary in the direction parallel to the magnetic field due to uni-directed currents of the CME.
We have confirmed the chiral magnetic skin effect in the model of Eq.(\ref{eq:BesWeyl}) \footnote{See Sec.\ref{sec:S4} in Supplemental Material.}.
Photonic systems \cite{Cerjan19,Guo09} and cold atoms \cite{Li19,Takasu20} may provide the spin-selective (or sublattice selective) loss term in Eq.(11), and thus the experimental realization of the chiral magnetic skin effect is feasible.


\smallskip
We are grateful to Masaya Nakagawa, Ken Shiozaki, Nobuyuki Okuma, Kohei Kawabata, Masaya Kunimi, and Taigen Kawano for valuable discussions. 
This work was supported by JST CREST Grant No.\ JPMJCR19T2, 
 KAKENHI Grant No.~JP20H00131, 
 and KAKENHI Grant No.~JP21J11810 from the JSPS.

\bibliographystyle{apsrev4-1}
\bibliography{DynamicalAnomaly3}

\widetext
\pagebreak

\renewcommand{\theequation}{S\arabic{equation}}
\renewcommand{\thefigure}{S\arabic{figure}}
\renewcommand{\thetable}{S\arabic{table}}
\renewcommand{\thesection}{S\arabic{section}}

\setcounter{equation}{0}
\setcounter{figure}{0}
\setcounter{table}{0}
\setcounter{section}{0}

\begin{center}
{\bf \large Supplemental Material}
\end{center}

\section{Gapless structures in non-Hermitian systems}
\label{sec:S0}

\begin{figure}[h]
 \centering
\includegraphics[width=90mm]{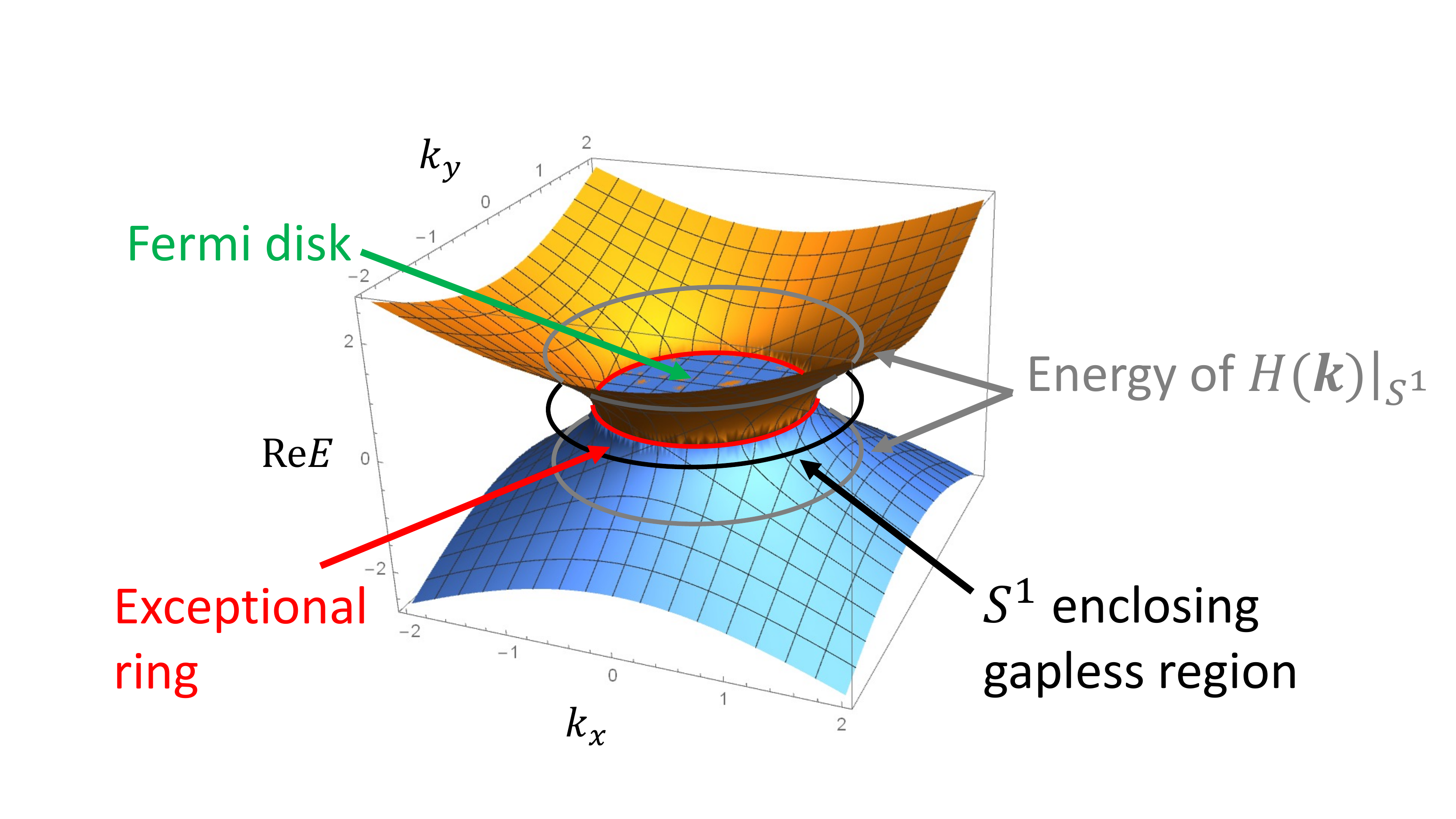}
\caption{A gapless structure in a two-dimensional non-Hermitian system. The gapless strucutre hosts an open Fermi disk and an exceptional ring, which are two-dimensional generalization of Fermi arc and exceptional point, respectively. There is a gap in the real part of the spectrum on $S^1$ enclosing the gapless region.}
\label{fig:S1}
\end{figure}

A non-Hermitian Hamiltonian $H({\bm k})$ exhibits an extended gapless structure in the momentum space.  
The gapless structure consists of an open region where the real part of the energy gap vanishes, and its boundary where the full complex energy gap vanishes and $H({\bm k})$ becomes defective. The open region and the boundary are called "Fermi arc" and "exceptional point" (or their higher dimensional generalization), respectively.  (See Fig.\ref{fig:S1}.)
When $H({\bm k})$ is deformed to be diagonalizable, the Fermi arc shrinks, the exceptional points are (pair-)annihilated, and the gapless structure reduces to a conventional Dirac or Weyl point. 
Thus, the topological charge of the Dirac or Weyl point protects the gapless structure.

In this paper, we focus on the gapless structures in AZ$^\dagger$ symmetry classes. Two different situations may happen: (i) the gapless structures are located at arbitrary positions in the complex energy plane, or (ii) they are located on the ${\rm Re}E=0$ axis of the complex energy plane.

The former situation occurs in classes A, AI$^\dagger$, and AII$^{\dagger}$. In these classes, one can multiply $H({\bm k})$ by a phase factor $e^{i\theta}$ and/or add a constant term to $H({\bm k})$ with keeping the symmetry classes and without closing a point gap, which changes the location of gapless structure.
On the other hand, the latter situation takes place
in classes AIII, BDI$^{\dagger}$, D$^{\dagger}$, DIII$^{\dagger}$, CII$^\dagger$, C$^{\dagger}$, and CI$^{\dagger}$: Because these classes support CS and/or PHS$^{\dagger}$, 
\begin{align}
\mbox{CS}: \Gamma H^{\dagger}({\bm k})\Gamma^{-1}=-H({\bm k}),
\quad
\mbox{PHS$^\dagger$}: C H^*({\bm k})C^{-1}=-H(-{\bm k}),
\end{align}
where $\Gamma$ and $C$ are unitary matrices, 
a complex energy band $E$ of $H$ is paired with $-E^*$.
Thus, gapless structures between these bands, which respect these symmetries, appear on the ${\rm Re}E=0$ axis.

As was mentioned above, the gapless structures support the topological charge of the Dirac or Weyl points.
We measure the topological charge as follows.
In case (i) in the above, we use a Fermi surface to define the topological charge. 
We introduce the Fermi surface $S$ as the set obeying $S=\{{\bm k}\in \mathbb{R}^d|{\rm Re}E_p({\bm k})=0\}$, where $E_p({\bm k})$ is a complex band energy of $H({\bm k})$.
Then, we define the topological charge by counting the Dirac or Weyl points inside $S$.
(See below for the precise definition.)
On the other hand, we define the topological charge in case (ii) as follows.
We encircle the whole of the gapless structure by a $(d-1)$-dimensional sphere $S^{d-1}$ in the momentum space, where $H({\bm k})$ is gapped in the real part of the energy spectrum. (See Fig.\ref{fig:S1}.)
From Theorem 2 in \cite{KSUS18},
the Hamiltonian $H({\bm k})$ on $S^{d-1}$ can be continuously deformed into a gapped Hermitian one, so
in a manner analogous to the Chern number for the conventional Weyl point, we define the topological charge on $S^{d-1}$, which is essentially identical to that in the Hermitian case.

When $H({\bm k})$ has a point gap, the above definition of the topological charge enables us to classify the gapless structures into two. In case (i), we classify them by using two possible Fermi surfaces: the Fermi surface with ${\rm Im}E_n(S)>0$ and that with ${\rm Im}E_n(S)<0$.
Gapless structures in case (i) provide nontrivial topological charges on these Fermi surfaces.
On the other hand, 
in case (ii), gapless structures satisfies ${\rm Re}E=0$. Thus in the presence of point gap, their energies are either positive in the imaginary part, ${\rm Im}E>0$, or negative in the imaginary part, ${\rm Im}E<0$.
Thus, we can classify them into two by using the sign of the imaginary part.



\section{PROOF OF EXTENDED NIELSEN-NINOMIYA THEOREM}
\label{sec:S2}

\begin{table}[b]
\caption{Extended Nielsen-Ninomiya theorem  for point gapped Hamiltonians in AZ$^{\dagger}$ symmetry classes for the spatial dimension $d\le 3$. 
In the first three AZ$^{\dagger}$ classes, 
gapless regions are located in an arbitrary position, and
in the other seven ones,
 gapless regions are located on the ${\rm Re}E=0$ axis.
The section numbers for the proof of Theorem are shown for each topological number.
}
\label{Table:S1}
\begin{tabular}{ccccc}
\hline\hline
AZ$^\dagger$ class & $d=0$ &$d=1$ &$d=2$ & $d=3$ \\
\hline
A & 0 & $\mathbb{Z}$ [Sec.\ref{sec:S2B1}]& 0 & $\mathbb{Z}$ [Sec.\ref{sec:S2B1}]\\
AI$^\dagger$ & 0 & 0 & 0 & $2\mathbb{Z}$ [Sec.\ref{sec:S2B1}]\\
AII$^{\dagger}$ & 0 & $\mathbb{Z}_2$ [\ref{sec:S2B3}] & $\mathbb{Z}_2$ [Sec.\ref{sec:S2B3}]& $\mathbb{Z}$ [Sec.\ref{sec:S2B1}]\\
\hline
AIII & $\mathbb{Z}$ [Sec.\ref{sec:S3A}]& 0 &  $\mathbb{Z}$ [Sec.\ref{sec:S2A}] & 0 \\
BDI$^{\dagger}$& $\mathbb{Z}$ [Sec.\ref{sec:S3B}]& 0 & 0 & 0 \\
D$^{\dagger}$ & $\mathbb{Z}_2$ [Sec.\ref{sec:S3C}]& $\mathbb{Z}$ [Sec.\ref{sec:S2A}]& 0 &0\\
DIII$^{\dagger}$ & $\mathbb{Z}_2$ [Sec.\ref{sec:S3D}]& $\mathbb{Z}_2$ [Sec.\ref{sec:S2A}]& $\mathbb{Z}$ [Sec.\ref{sec:S2A}]& 0\\
CII$^{\dagger}$ & 2$\mathbb{Z}$ [Sec.\ref{sec:S3E}]& 0 & $\mathbb{Z}_2$ [Sec.\ref{sec:S2A}]& $\mathbb{Z}_2$ [Sec.\ref{sec:S2A}]\\
C$^{\dagger}$ & 0 &2$\mathbb{Z}$ [Sec.\ref{sec:S2A}]& 0 & $\mathbb{Z}_2$ [Sec.\ref{sec:S2A}]\\
CI$^{\dagger}$ & 0 & 0 & $2\mathbb{Z}$ [Sec.\ref{sec:S2A}]& 0\\
\hline\hline
\end{tabular} 
\end{table}

In this section, we prove the following theorem.

\vspace{3ex}
\noindent
{\bf Theorem } 
Let $H({\bm k})$ be a Hamiltonian with a point gap ($\det (H-E_{\rm F})\neq 0$) in an AZ$^{\dagger}$ class.
Then, gapless structures in $H({\bm k})$ obey the following relation,
\begin{align}\label{eq:SThm1}
n=\sum_{\text{Im}E_\alpha>0}\nu_{\alpha}
=-\sum_{\text{Im}E_\alpha<0}\nu_{\alpha},
\end{align}
where $n$ is the bulk topological invariant of the point gapped Hamiltonian $H({\bm k})$.
In case (i) of Sec.~\ref{sec:S0}, $\nu_\alpha$ is the topological charge on the $\alpha$-th Fermi surface (See Sec.\ref{sec:S0}), and $E_\alpha$ is the complex energy of the $\alpha$-th Fermi surface measured from $E_{\rm F}$.
In case (ii) of Sec.~\ref{sec:S0}, $\nu_\alpha$ is the topological charge of the $\alpha$-th gapless structure, and $E_\alpha$ is the complex energy of the $\alpha$-th gapless structures measured from $E_{\rm F}$.
Here the reference energy $E_{\rm F}$ for a point gap should be invariant under symmetry.

\vspace{3ex}
For class A in one- and three- dimensions, the above theorem gives Theorem 1 and Theorem 2 in the main text, respectively.

\subsection{Case (i)}
\label{sec:S2B}

In case (i) (classes A, AI$^\dagger$ and AII$^{\dagger}$), we prove Theorem by direct evaluation of $n$.
Below, we assume without loss of generality that $E_{\rm F}=0$.
If necessary, we recover $E_{\rm F}$ by replacing $H({\bm k})$ with $H({\bm k})-E_{\rm F}$.

\subsubsection{Class A}
\label{sec:S2B1}

First, we consider a class A non-Hermitian Hamiltonian $H({\bm k})$.
The point gap topological number $n$ in $d=2q+1$ dimensions ($q=0,1,\dots$) 
is the winding number $w_{2q+1}$,
\begin{align}
n=w_{2q+1}=
\left(\frac{i}{2\pi}\right)^{q+1}
\frac{q!}{(2q+1)!}
\int_{\rm BZ} {\rm tr}[H^{-1}{\rm d}H]^{2q+1}.
\label{eq:Swinding}
\end{align}
To prove the Theorem,  we use the technique developed in Refs.~\cite{Sato09, QTZ10, STYY11}. 
We first deform the Hamiltonian $H({\bm k})$ into a unitary matrix, which is possible with keeping a point gap \cite{Gong18,KSUS18}.
As $H({\bm k})$ remains invertible during this deformation, this procedure does not change $n$.
After this, $H({\bm k})$ is diagonalizable and can be written as
\begin{align}
H({\bm k})=\sum_{p}E_p({\bm k})|u_p({\bm k}\rangle\langle u_p({\bm k})|,
\quad |E_p({\bm k})|=1,
\label{eq:SUH}
\end{align}
where $|u_p({\bm k})\rangle$ is an eigenstate of $H({\bm k})$ with an eigenvalue $E_p({\bm k})$.
We furthermore deform $H({\bm k})$ as follows,
\begin{align}
H({\bm k})=\sum_{p}
e^{i\theta_p({\bm k})}
|u_p({\bm k}\rangle\langle u_p({\bm k})|,
\label{eq:thetaH}
\end{align}
with
\begin{align}
e^{i\theta_p({\bm k})}=
\frac{{\rm Re}E_p({\bm k})+\lambda i{\rm Im}E_p({\bm k})}
{|{\rm Re}E_p({\bm k})+\lambda i{\rm Im}E_p({\bm k})|}
\label{eq:SUH2}
\end{align}
where $0<\lambda\le 1$ is a deformation parameter. 
When $\lambda=1$, $H({\bm k})$ returns to Eq.~(\ref{eq:SUH}).
As $|E_p({\bm k})|\neq 0$, this Hamiltonian is also invertible as long as $\lambda\neq 0$, and thus has the same value of $n$.
Now take the limit $\lambda\rightarrow 0$, where $\lambda$ is infinitesimally tiny but nonzero.
In this limit, the eigenvalue $e^{i\theta_p({\bm k})}$ takes a constant value $e^{i\theta_p({\bm k})}=\pm 1$ except on the Fermi surfaces $S_{p\alpha}$ defined by $\{{\bm k}\in S_{p\alpha}|{\rm Re}E_{p}({\bm k})=0\}$. 
The Fermi surface generally consists of a set of connected components, and $\alpha$ labels each connected component of the Fermi surface.
Near the Fermi surfaces, $\theta_p({\bm k})$ satisfies 
\begin{align}
\nabla_{\bm k}
\theta_p({\bm k})=-\pi {\rm sgn}[{\rm Im}E_p({\bm k})]\delta({\rm ReE}_p({\bm k}))\nabla_{\bm k}[{\rm Re}E_p({\bm k})],
\label{eq:theta}
\end{align}
from which the branch of $\theta_p({\bm k})$ is determined uniquely.
${\rm Im}E_p({\bm k})$ takes the same sign on each connected component $S_{p\alpha}$
since $|E_p({\bm k})|=1$,  and thus ${\rm sgn}[{\rm Im}E_{p}({\bm k}))]$ in Eq.~(\ref{eq:theta}) is well-defined. Substituting Eq.~(\ref{eq:thetaH}) with a tiny $\lambda>0$, we obtain Theorem.

For instance, let us consider the $d=1$ ($q=0$) case, where $n$ is given by the 1D winding number 
\begin{align}
w_1&=-\frac{1}{2\pi i}\int_{-\pi}^{\pi} dk {\rm tr}[H^{-1}(k)\partial_k H(k)]
\nonumber\\
&=-\frac{1}{2\pi i} \int_{-\pi}^{\pi}dk \partial_k \ln {\rm det}H(k).
\end{align} 
Substituting Eq.~(\ref{eq:thetaH}) into this, we obtain
\begin{align}
w_1&=-\frac{1}{2\pi}\int_{-\pi}^{\pi} dk \sum_p\partial_k\theta_p(k)
\nonumber\\
&=\frac{1}{2}\sum_{p\alpha}\int_{-\pi}^{\pi}dk\, 
{\rm sgn}[{\rm Im}E_p(k_{p\alpha})]\delta(k-k_{p\alpha}) {\rm sgn}[\partial_k [{\rm Re}E_p(k_{p\alpha})]]
\nonumber\\
&=\frac{1}{2}\sum_{p\alpha} {\rm sgn}[{\rm Im}E_p(k_{p\alpha})]{\rm sgn}[\partial_k [{\rm Re}E_p(k_{p\alpha})]],
\label{eq:localization0}
\end{align}
where $k_{p\alpha}$ is the Fermi point defined by ${\rm Re}E_p(k_{p\alpha})=0$.
Now we use the original Nielsen-Ninomiya (NN) theorem. 
Since the real part of $H({\bm k})$
\begin{align}
{\rm Re}H({\bm k})=\sum_p {\rm Re}E_p({\bm k})|u_p({\bm k})\rangle\langle u_p({\bm k})|
\end{align}
is Hermitian, it obeys the original NN theorem.
For $d=1$, the NN theorem yields that
\begin{align}
\sum_{p\alpha}{\rm sgn}[\partial_k [{\rm Re}E_p(k_{p\alpha})]]=0,    
\end{align}
which is equivalent to
\begin{align}
\sum_{{\rm Im}E_p(k_{p\alpha})>0}{\rm sgn}[\partial_k [{\rm Re}E_p(k_{p\alpha})]]        
=-\sum_{{\rm Im}E_p(k_{n\alpha})<0}{\rm sgn}[\partial_k [{\rm Re}E_p(k_{p\alpha})]].        
\end{align}
Therefore, from Eq.~(\ref{eq:localization0}), we obtain Theorem for $d=1$ (Theorem 1 in the main text),
\begin{align}
w_1=\sum_{{\rm Im}E_{p}(k_{p\alpha})>0} {\rm sgn}[\partial_k [{\rm Re}E_p(k_{p\alpha})]]    
=-\sum_{{\rm Im}E_{p}(k_{p\alpha})<0} {\rm sgn}[\partial_k [{\rm Re}E_p(k_{p\alpha})]].    
\end{align}

In a similar manner, we can also derive Theorem for $d=3$ ($q=1$).
In this case, $n$ is given by the 3D winding number, which is evaluated as \cite{QTZ10}
\begin{align}
w_3=\frac{1}{2}\sum_{p\alpha}{\rm sgn}[{\rm Im} E_p(S_{p\alpha})]{\rm Ch}(S_{p\alpha}), 
\label{eq:localization}
\end{align}
where ${\rm Ch}(S_{p\alpha})$ is the Chern number on $S_{p\alpha}$ defined by
\begin{align}
{\rm Ch}(S_{p\alpha})=\frac{1}{2\pi i}
\int_{S_{p\alpha}} (\nabla\times {\bm A}_p)\cdot d{\bm S}.
\end{align}
Here ${\bm A}_p=\langle u_p ({\bm k}) |\nabla u_p({\bm k})\rangle$ is the connection of the eigenstate $|u_p\rangle$ for ${\rm Re}H({\bm k})$, 
\begin{align}
{\rm Re}H({\bm k})|u_p({\bm k})\rangle={\rm Re}E_p({\bm k})|u_p({\bm k})\rangle,      
\end{align}
and the orientation of $S_{p\alpha}$ is chosen as the direction of the Fermi velocity $\partial_{\bm k}[{\rm Re}E_p({\bm k})]_{{\bm k}\in S_{p\alpha}}$.
To obtain Theorem from Eq.~(\ref{eq:localization}), we again use the original NN theorem for ${\rm Re}H({\bm k})$.
As we shall argue in Sec.~\ref{sec:S3}, the NN theorem yields 
\begin{align}
\sum_{p\alpha}{\rm Ch}(S_{p\alpha})=0,
\end{align}
which is recast into
\begin{align}
\sum_{{\rm Im}E_p(S_{p\alpha})>0}
{\rm Ch}(S_{p\alpha})
=-\sum_{{\rm Im}E_p(S_{p\alpha})<0}
{\rm Ch}(S_{p\alpha}).
\end{align}
Using this relation, we finally obtain Theorem for $d=3$ (Theorem 2 in the main text),
\begin{align}
w_3=\sum_{{\rm Im}E_p(S_{p\alpha})>0}
{\rm Ch}(S_{p\alpha})
=-\sum_{{\rm Im}E_p(S_{p\alpha})<0}
{\rm Ch}(S_{p\alpha}).
\label{eq:Thm1classA}
\end{align}

We can prove the Theorem for general $d$ in the same manner.
For ${\bm k}$ other than on the Fermi surfaces, we have $H^{-1}({\bm k})=H({\bm k})$, and thus ${\rm tr}[H^{-1} dH]^{2q+1}=0$. 
Therefore, only the Fermi surfaces contribute to the integral of Eq.(\ref{eq:Swinding}).
Furthermore, a straightforward calculation shows that only terms with odd powers of $d\theta_p$ are nonzero in ${\rm tr}[H^{-1} dH]^{2q+1}$, 
and thus from Eq.(\ref{eq:theta}), the Fermi surface contribution of the integral has the form of 
\begin{align}
n=\sum_p {\rm sgn}[{\rm Im} E_{p}] \eta_p,    
\label{eq:eta}
\end{align}
where $\eta_p$ is an integral on the Fermi surface.
(Here we have used that each Fermi surface has a definite sign of ${\rm Im} E_p$ since $H({\bm k})$ has a point gap.)
Similarly, we also have
\begin{align}
\sum_p \eta_p=0.    
\end{align}
by considering $n$ for ${\rm Re}H({\bm k})+i\delta$  ($\delta>0$).
Thus, Eq.(\ref{eq:eta}) yields
\begin{align}
n=\sum_{{\rm Im}E_p>0} 2\eta_p =-\sum_{{\rm Im}E_p<0} 2\eta_p, \end{align}
which leads to Theorem with $\nu_\alpha=2\eta_p$.
Note that $\nu_p$ is an integer since $n$ is an integer, and $\nu_p$ is zero
if there is no gapless point inside the Fermi surface as the Fermi surface can shrink and vanish smoothly in this case. Therefore, a non-zero $\nu_p$ implies gapless points inside the Fermi surface.


\subsubsection{Class AI$^{\dagger}$}
\label{sec:S2B2}

A point gapped Hamiltonian in class AI$^{\dagger}$ has the $2\mathbb{Z}$ index in three dimensions.
(See Table.\ref{Table:S1}.)
The $2\mathbb{Z}$ invariant $n$ is given by the winding number in Eq.(\ref{eq:Swinding}) with $q=1$, and thus
we obtain Theorem in the form of Eq.(\ref{eq:Thm1classA}) in the same manner to class A with $d=3$.

\subsubsection{Class AII$^{\dagger}$}
\label{sec:S2B3}

A Hamiltonian $H({\bm k})$ in class AII$^{\dagger}$ satisfies 
\begin{align}
TH^T({\bm k})T^{\dagger}=H(-{\bm k}), 
\label{seq:TRS}
\end{align}
where $T$ is a unitary matrix with $TT^*=-1$.
For point gap topological numbers in class AII$^{\dagger}$,  
we have the $\mathbb{Z}_2$ invariants in one and two-dimensions and the $\mathbb{Z}$ invariant in three dimensions.
In three dimensions, the $\mathbb{Z}$ invariant is given by the winding number in Eq.(\ref{eq:Swinding}) with $q=1$ again, so Theorem holds in the same manner to class A with $d=3$.
Therefore, we prove below the Theorem in one and two dimensions.

First, we consider the one-dimensional case. In one dimension, the $\mathbb{Z}_2$ invariant is given by
\begin{align}
(-1)^{n}={\rm sgn}\left\{
\frac{{\rm Pf}[H(\pi)T]}{{\rm Pf}[H(0)T]}{\rm exp}\left[
-\frac{1}{2}\int_{k=0}^{k=\pi}dk \partial_k {\rm log}{\rm det}[H(k)T]
\right]
\right\}.
\label{seq:z21d}
\end{align}
After the unitarization, $H(k)$ is recast into
\begin{align}
H(k)&=\sum_p e^{i\theta_p(k)}|u_p(k)\rangle \langle u_p(k)|
\nonumber\\
&=
U(k)\left(
\begin{array}{ccc}
e^{i\theta_1(k)} & & \\
& e^{i\theta_2(k)} &\\
& & \ddots
\end{array}
\right)
U^{\dagger}(k)
\end{align}
with the unitary matrix $U({\bm k})=\left(|u_1(k)\rangle, |u_2(k)\rangle, \cdots \right)$, so we have
\begin{align}
{\rm det}H(k)={\rm exp}(i\sum_p \theta_p(k)). 
\end{align}
Therefore, Eq.(\ref{eq:theta}) leads to
\begin{align}
\int_0^\pi dk \partial_k {\rm log}{\rm det}[H(k)T]&=
\int_0^\pi dk \partial_k\left[i\sum_p\theta_p(k)\right]
\nonumber\\
&=-\pi \sum_{p\alpha}{\rm sgn}\left[
\partial_k {\rm Re}E_p(k_{p\alpha})\,
{\rm Im}E_p(k_{p\alpha})\right],
\end{align}
where $0<k_{p\alpha}<\pi$ is the $\alpha$-th Fermi point defined by  ${\rm Re}E_p(k_{p\alpha})=0$, 
and we have assumed that any Fermi point does not exist at the time-reversal invariant momentum $k=0,\pi$ without loss of generality.
The exponential term in Eq.(\ref{seq:z21d}) is evaluated as
\begin{align}
{\rm exp}\left[-\frac{1}{2}\int_0^\pi dk \partial_k {\rm log}{\rm det}[H(k)T]\right] 
=\prod_{n\alpha}i{\rm sgn}\left[\partial_k {\rm Re}E_p(k_{p\alpha})\right]
\prod_{p\alpha}{\rm sgn}\left[{\rm Im}E_p(k_{p\alpha})\right].
\label{eq:z2exp}
\end{align}
Here we note that each Fermi point with positive (negative) $\partial_k {\rm Re}E_p(k_{p\alpha})$ decreases (increases) the number of the occupied state $N_{\rm occ}(0)$ ($N_{\rm occ}(\pi)$) at $k=0$ ($k=\pi$) by 1, where the occupied state is defined as a state with negative ${\rm Re}E_p(k)$.
Thus, the first product of the right-hand side in Eq.(\ref{eq:z2exp}) becomes
\begin{align}
\prod_{p\alpha}i{\rm sgn}\left[\partial_k {\rm Re}E_p(k_{p\alpha})\right]=
i^{N_{\rm occ}(0)-N_{\rm occ}(\pi)}
=(-1)^{[N_{\rm occ}(0)-N_{\rm occ}(\pi)]/2},
\end{align}
which leads to
\begin{align}
{\rm exp}\left[-\frac{1}{2}\int_0^\pi dk \partial_k {\rm log}{\rm det}[H(k)T]\right] 
=(-1)^{[N_{\rm occ}(0)-N_{\rm occ}(\pi)]/2} 
\prod_{p\alpha}{\rm sgn}\left[{\rm Im}E_p(k_{p\alpha})\right].
\label{seq:exp2}
\end{align}
It should be noted here that $N_{\rm occ}(k)$ ($k=0,\pi$) is an even number because of the generalized Kramers degeneracy in class AII$^{\dagger}$ \cite{SHEK12,ZL2019,KSUS18}.

\begin{figure}[h]
 \centering
\includegraphics[width=55mm]{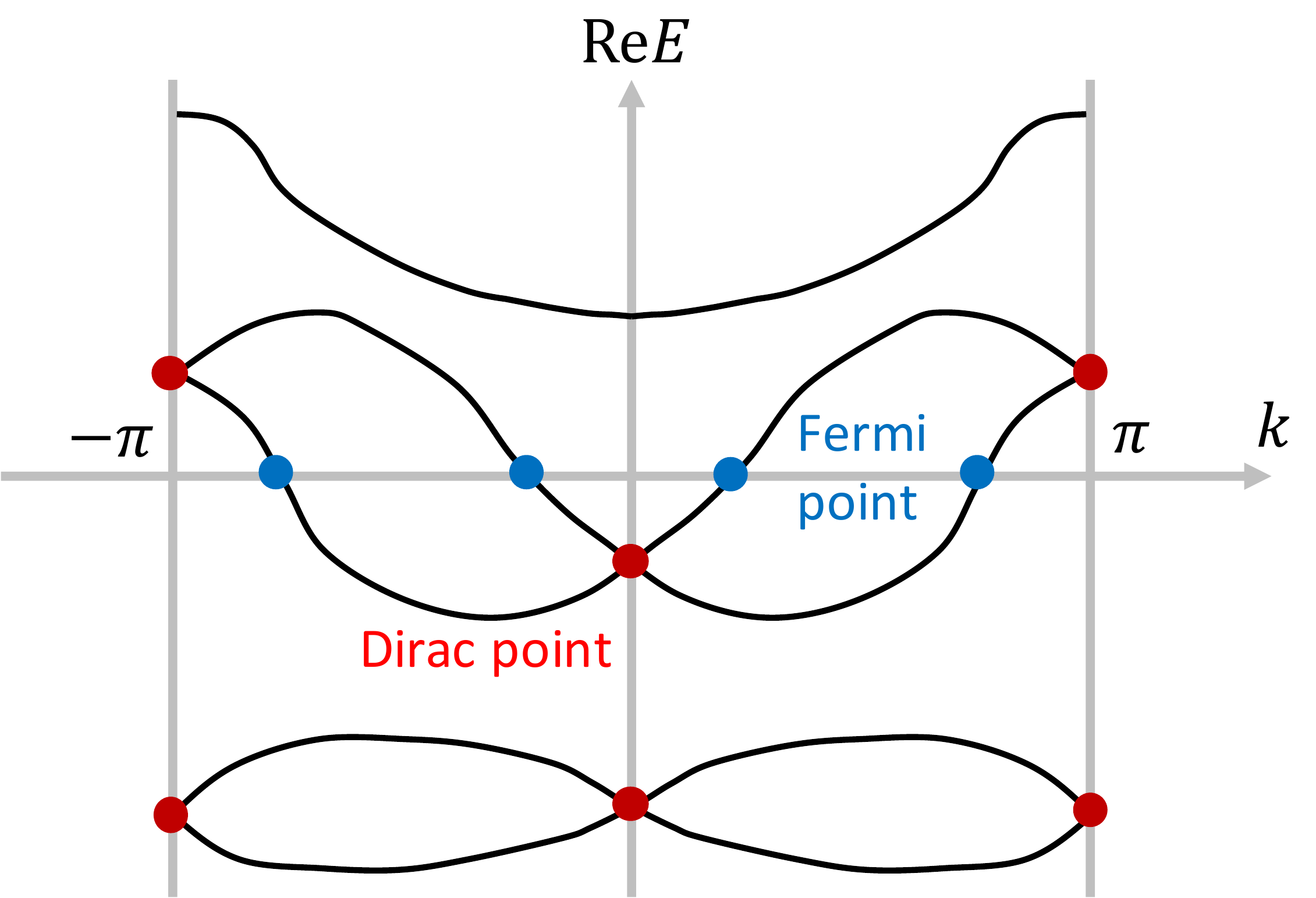}
\caption{Dirac and Fermi points in a one-dimensional class AII system. Each Kramers pair of Fermi points encloses a Dirac point at a time-reversal invariant momentum.}
\label{fig:S2}
\end{figure}

We now evaluate the Paffians in Eq.(\ref{seq:z21d}) as follows. 
At the time-reversal invariant momentum $k_0=0,\pi$, $\theta_p(k_0)$ takes either $0$ or $\pm \pi$, so $H(k_0)$ is
\begin{align}
H(k_0)=U(k_0)\Lambda(k_0)U^{\dagger}(k_0), \quad (k_0=0,\pi)
\end{align}
with 
\begin{align}
\Lambda(k_0)=\left(
\begin{array}{cc}
1_{N_{\rm emp}(k_0)\times N_{\rm emp}(k_0)} & \\
& -1_{N_{\rm occ}(k_0)\times N_{\rm occ}(k_0)} \\
\end{array}
\right).
\end{align}
Therefore, the Paffian becomes
\begin{align}
{\rm Pf}\left[H(k_0)T\right]
&={\rm Pf}\left[U(k_0)\Lambda(k_0)U^{\dagger}(k_0)T\right] 
\nonumber\\
&={\rm Pf}\left[U^\dagger (k_0)U(k_0)\Lambda(k_0)U^{\dagger}(k_0)TU^*(k_0)\right]/{\rm det}U^*(k_0)
\nonumber\\ 
&={\rm Pf}\left[\Lambda(k_0)U^{\dagger}(k_0)TU^*(k_0)\right]{\rm det}U(k_0),
\end{align}
where we have used the formula ${\rm Pf}[B^T A B]={\rm Pf}[A]{\rm det}B$ for an antisymmetric matrix $A$.
Because of TRS$^\dagger$ in Eq.(\ref{seq:TRS}), it holds that
\begin{align}
\left[U^\dagger(k_0)T U^*(k_0),\Lambda(k_0)\right]=0, 
\end{align}
so
$U^{\dagger}(k_0)TU^*(k_0)$ are block-diagonal.
Hence, we have
\begin{align}
{\rm Pf}\left[\Lambda(k_0)U^\dagger (k_0)T U^*(k_0)\right]
&=(-1)^{N_{\rm occ}(k_0)/2}
{\rm Pf}\left[U^\dagger (k_0)T U^*(k_0)\right] 
\nonumber\\
&=(-1)^{N_{\rm occ}(k_0)/2}
{\rm Pf}[T] {\rm det}U^*(k_0),
\end{align}
which implies 
\begin{align}
{\rm Pf}\left[H(k_0)T\right]=(-1)^{N_{\rm occ}(k_0)/2}{\rm Pf}[T].
\end{align}
Substituting the above result and Eq.(\ref{seq:exp2}) into the right-hand side of Eq.(\ref{seq:z21d}), we obtain 
\begin{align}
(-1)^{n}=\prod_{p\alpha}{\rm sgn}[{\rm Im}E_p(k_{p\alpha})]
=(-1)^{\sum_{{\rm Im}E_p(k_{p\alpha})>0}}.
\end{align}
Because of the Kramers degeneracy at $k=0,\pi$, the total number of the Fermi points between $k=0$ and $k=\pi$ is even, which implies $
(-1)^{\sum_{{\rm Im}E_p(k_{p\alpha})>0}+\sum_{{\rm Im}E_p(k_{p\alpha})<0}}=1$.
Therefore, we have
\begin{align}
(-1)^{n}=(-1)^{\sum_{{\rm Im}E_p(k_{p\alpha})>0}}
=(-1)^{\sum_{{\rm Im}E_p(k_{p\alpha})<0}},
\end{align}
which gives Eq.(\ref{eq:SThm1}) in Theorem for class AII$^{\dagger}$ in one-dimension:
\begin{align}
n= \sum_{{\rm Im}E_p(k_{p\alpha})>0}=-\sum_{{\rm Im}E_p(k_{p\alpha})>0}
\quad (\mbox{mod.2})
\label{seq:Th1AII1d}
\end{align}
As we illustrate in Fig.\ref{fig:S2}, each Kramers pair of Fermi points always encloses a gapless Dirac point at time-reversal invariant momentum because of the Kramers degeneracy.
Therefore, the middle and the right-hand side in Eq.(\ref{seq:Th1AII1d}) counts the Dirac points enclosed by the Fermi point. 


Next, we consider the two-dimensional case. 
The $\mathbb{Z}_2$ invariant $(-1)^n$ in two dimensions is given as the product of the one-dimensional $\mathbb{Z}_2$ invariants,
\begin{align}
(-1)^{n}={\rm sgn}\left\{
\frac{{\rm Pf}[H(\pi,0)T]}{{\rm Pf}[H(0,0)T]}{\rm exp}\left[
-\frac{1}{2}\int_{k_x=0}^{k_x=\pi}dk_x \partial_{k_x} {\rm log}{\rm det}[H(k_x,0)T]
\right]
\right\}
\nonumber\\
\times
{\rm sgn}\left\{
\frac{{\rm Pf}[H(\pi,\pi)T]}{{\rm Pf}[H(0,\pi)T]}{\rm exp}\left[
-\frac{1}{2}\int_{k_x=0}^{k_x=\pi}dk_x \partial_{k_x} {\rm log}{\rm det}[H(k_x,\pi)T]
\right]
\right\}.
\label{seq:z22d}
\end{align}
Thus, like the above, we obtain 
\begin{align}
n= 
\sum_{{\rm Im}E_p(k_{p\alpha},0)>0}+\sum_{{\rm Im}E_p(k'_{p\alpha},\pi)>0}
=-
\left(
\sum_{{\rm Im}E_p(k_{p\alpha},0)<0}+\sum_{{\rm Im}E_p(k'_{p\alpha},\pi)<0}
\right)
\quad (\mbox{mod.2}),
\end{align}
where $0<k_{p\alpha}<\pi$ ($0<k'_{p\alpha}<\pi$) is the $\alpha$-th Fermi point defined by ${\rm Re}E_p(k_{p\alpha},0)=0$ (${\rm Re}E_{p}(k'_{p\alpha},\pi)=0$). 
Note that the Fermi points in the above are the intersection between the Fermi surface $S_{p\alpha}$ defined by ${\rm Re}E_p(S_{p\alpha})=0$ and the $k_y=0, \pi$ lines. 
We also notice that since each Fermi point and its Kramers partner enclose a time-reversal invariant momentum, the summation
\begin{align}
\sum_{{\rm Im}E_p(k_{p\alpha},0)>0}+\sum_{{\rm Im}E_p(k'_{p\alpha},\pi)>0} 
\end{align}
counts the number of time-reversal invariant momenta enclosed by the Fermi surfaces with positive ${\rm Im}E_p(S_{p\alpha})$. 
Thus, we can rewrite it as
\begin{align}
\sum_{{\rm Im}E_p(k_{p\alpha},0)>0}+\sum_{{\rm Im}E_p(k'_{p\alpha},\pi)>0}=\sum_{{\rm Im}E_p(S_{p\alpha})>0}m_{p\alpha}, 
\end{align}
where $m_{p\alpha}$ is the number of time-reversal invariant momenta enclosed by the Fermi surface $S_{p\alpha}$.
Consequently, we obtain Eq.(\ref{eq:SThm1}) in Theorem for class AII$^{\dagger}$ in two dimensions,
\begin{align}
n= 
\sum_{{\rm Im}E_p(S_{p\alpha})>0}m_{p\alpha}
=
\sum_{{\rm Im}E_p(S_{p\alpha})<0}m_{p\alpha}, 
\quad (\mbox{mod.2}).
\label{seq:Th1AII2d}
\end{align}
Again, since a Fermi surface enclosing a time-reversal invariant momentum also encloses a Dirac point at the time-reversal invariant point, the second and the third terms in Eq.(\ref{seq:Th1AII2d}) count the number of Dirac points inside the Fermi surfaces. 
\subsection{Case (ii)}
\label{sec:S2A}
We here provide proof of Eq.~(\ref{eq:SThm1}) in case (ii) of Sec.~\ref{sec:S0}.
The key idea is to use a primitive model that generates all relevant topological states.
It has been known that by stacking the generator via a direct sum and performing a smooth deformation, any point gapped Hamiltonian 
can be produced (up to addition/subtraction of trivial bands). 
Therefore, Eq.~(\ref{eq:SThm1}) can be proved by proving it for the generator.
We choose such a primitive model with $E_{\rm P}=0$ as follows \cite{Lee19}, 
\begin{align}
H(\bm{k})=h(\bm{k})+i \gamma(\bm{k}),
\label{eq:primitive}
\end{align}
with
\begin{align}
h(\bm{k})=\sum_{j=1}^d \sin k_{j} \Gamma_{j}, \quad
\gamma(\bm{k})=m+\sum_{j=1}^d \cos k_{j},
\label{eq:hg}
\end{align}
where the gamma matrix $\Gamma_i$ is Hermitian and obeys $\{\Gamma_i,\Gamma_j\}=2\delta_{ij}$, CS and PHS$^{\dagger}$ imply $\{\Gamma, \Gamma_i\}=0$ and $C \Gamma_i^*+\Gamma_i C=0$, respectively. We also assume that $m$ is a real parameter with $-d<m<-d+2$.
Note that $\gamma({\bm k})$ in Eq.~(\ref{eq:hg}) is consistent with any symmetry of AZ$^{\dagger}$ classes.
The energy spectrum of this model is
\begin{align}
E({\bm k})=\pm \sqrt{\sum_{j=1}^d \sin^2 k_j}+i(m+\sum_{j=1}^d \cos k_j).
\end{align}
At any time-reversal invariant momenta, this model exhibits a gapless region in the form of a Dirac/Weyl point.
Because of CS and/or PHS$^{\dagger}$ in case (ii), all the Dirac/Weyl points are located on the ${\rm Re}E=0$ axis in the complex energy plane.  

Now we prove the Theorem for the generator.
When $-d< m <-d+2$, only a single Dirac/Weyl point with ${\bm k}={\bm 0}$ has a positive ${\rm Im}E({\bm k})$. 
It is described by $H({\bm k})=\sum_{j=1}^d k_j\Gamma_j+i(m+d)$ near ${\bm k}=0$, so its topological charge is $\nu=1$.
All other $2^d-1$ Dirac/Weyl points have negative ${\rm Im}E({\bm k})$s', and their total topological charge are $\nu=-1$. 
Therefore, the second equality of Eq.~(\ref{eq:SThm1}) holds.
We can also show that $n=1$ as follows.
The topological number $n$ is given as the topological number of the following Hermitian Hamiltonian,
\begin{eqnarray}
\widetilde{H}(\bm{k})=
\pmat{ & H(\bm{k})\\
H^\dagger(\bm{k}) & }=
\tau_{x} \otimes h(\bm{k})-\tau_{y} \otimes \gamma(\bm{k}).
\end{eqnarray}
For $m<-d$, $\gamma({\bm k})$ is always negative in the whole region of ${\bm k}$, implying $n=0$. 
Then, as one increases $m$, the point gap closes (i.e. ${\rm det}H({\bm k})=0$) at $m=-d$, which provides a (simple) topological phase transition of $\widetilde{H}({\bm k})$. 
As a result, in the parameter region $-d<m<-d+2$, we have $n=1$. 
Therefore, the first equality of Eq.~(\ref{eq:SThm1}) also holds.

\section{Extended Nielsen-Ninomiya Theorem in zero dimension}

This section shows how Theorem in Sec.\ref{sec:S2} works in zero dimension, where the Hamiltonian $H$ is merely a matrix. Below, we assume $E_{\rm F}=0$ without loss of generality.

\subsection{Class AIII}
\label{sec:S3A}

In class AIII, the Hamiltonian obeys 
\begin{align}
\Gamma H^{\dagger}\Gamma^{-1}=-H, \quad \Gamma^2=1, 
\label{seq:AIII0d}
\end{align}
with a unitary matrix $\Gamma$.
Since $iH\Gamma$ is hermitian, it is diagonalizable and has real eigenvalues.
The point gap topological number $n$ is defined as
\begin{align}
n=\frac{1}{2}\left[N_-(iH\Gamma)-N_+(iH\Gamma)\right], 
\label{seq:npAIII0d}
\end{align}
where $N_{-}(iH\Gamma)$ ($N_{+}(iH\Gamma)$) is the number of negative (positive) eigenstates in $iH\Gamma$.
We also introduce a topological number $\nu$ for gapless states as follows.
Consider the right and left eigenstates of $H$ with the eigenvalue $E_p$,
\begin{align}
H|u_p\rangle=E_p|u_p\rangle, \quad
\langle\!\langle u_p| H
=E_p\langle\!\langle u_p|.
\end{align}
Then, CS implies 
\begin{align}
H[\Gamma |u_p\rangle\!\rangle]&=-\Gamma H^{\dagger}|u_p\rangle\!\rangle
\nonumber\\
&=-E_p^* [\Gamma|u_p\rangle\!\rangle].
\end{align}
Thus, when the eigenstates are gapless in the real part of the eigenvalue, {\it i.e.} ${\rm Re}E_p=0$, 
$\Gamma |u_p\rangle\!\rangle$ and $|u_p\rangle$ have the same eigenvalue.
In this case, with a suitable choice of the biorthogonal basis, they satisfy
\begin{align}
\Gamma|u_p\rangle\!\rangle=\pm |u_p\rangle. 
\end{align}
The sign on the right-hand side is the non-Hermitian version of the chirality, which defines the topological number $\nu_p$ of the gapless state $|u_p\rangle$ as $\nu_p=\pm 1$ \cite{Esaki11}.


To check Theorem, we consider a general $2\times 2$ matrix in class AIII,
\begin{align}
H=ia_0\sigma_0+a_1\sigma_1+a_2\sigma_2+ia_3\sigma_3, \quad \Gamma=\sigma_3, 
\label{seq:2x2AIII}
\end{align}
where $a_\mu$ are real parameters. The eigenvalus of $H$ are given by
\begin{align}
E=ia_0\pm \sqrt{a_1^2+a_2^2-a_3^2}\equiv E_{\pm}, 
\end{align}
and thus $H$ supports gapless states defined by ${\rm Re}E_\pm=0$ when $a_3^2>a_1^2+a_2^2$.
The corresponding biorthogonal eigenstates are given by
\begin{align}
|\pm\rangle =
\frac{1}{\sqrt{2\sqrt{a_3^2-(a_1^2+a_2^2)}\left|a_3\pm\sqrt{a_3^2-(a_1^2+a_2^2)}\right|}}
\left(
\begin{array}{c}
ia_3\pm i \sqrt{a_3^2-(a_1^2+a_2^2)} \\
a_1+ia_2
\end{array}
\right),
\nonumber\\
|\pm\rangle\!\rangle =
\frac{\pm {\rm sgn}[a_3\pm \sqrt{a_3^2-(a_1^2+a_2^2)}]}
{\sqrt{2\sqrt{a_3^2-(a_1^2+a_2^2)}\left|a_3\pm \sqrt{a_3^2-(a_1^2+a_2^2)}\right|}}
\left(
\begin{array}{c}
ia_3\pm i \sqrt{a_3^2-(a_1^2+a_2^2)} \\
-a_1-ia_2
\end{array}
\right),
\end{align}
where $|\pm \rangle$ and $|\pm \rangle\!\rangle$ are right and left eigenstates with the eigenvalues $E_{\pm}$.
These gapless states satisfy
\begin{align}
\sigma_3 |\pm\rangle\!\rangle=\pm {\rm sgn}
\left[a_3\pm\sqrt{a_3^2-(a_1^2+a_2^2)}\right]|\pm\rangle,  
\label{seq:chirality2x2}
\end{align} 
so $|+\rangle$ and $|-\rangle$ have the topological numbers,
\begin{align}
\nu_{\pm}=\pm {\rm sgn}\left[a_3\pm \sqrt{a_3^2-(a_1^2+a_2^2)}\right]. 
\end{align}
Below we evaluate 
\begin{align}
\sum_{{\rm Im}E_p>0}\nu_p, \quad \sum_{{\rm Im}E_p<0}\nu_p,	 	 
\end{align}
for the gapless states. We consider four different cases, 
(i) $a_3>\sqrt{a_0^2+a_1^2+a_2^2}$,
(ii) $\sqrt{a_0^2+a_1^2+a_2^2}>a_3>\sqrt{a_1^2+a_2^2}$, or $-\sqrt{a_1^2+a_2^2}>a_3>-\sqrt{a_0^2+a_1^2+a_2^2}$, 
(iii) $\sqrt{a_1^2+a_2^2}>a_3>-\sqrt{a_1^2+a_2^2}$, 
(iv) $-\sqrt{a_0^2+a_1^2+a_2^2}>a_3$. 

\subsubsection{Case with $a_3>\sqrt{a_0^2+a_1^2+a_2^2}$}

Because $a_3>\sqrt{a_1^2+a_2^2}$, it holds that ${\rm sgn}[a_3\pm\sqrt{a_3^2-(a_1^2+a_2^2)}]=1$. Thus
Eq.(\ref{seq:chirality2x2}) implies that the gapless state $|\pm \rangle$ has $\nu_{\pm}=\pm 1$.
Noting that ${\rm sgn}\left[{\rm Im}E_{\pm}\right]={\rm sgn}[a_0\pm \sqrt{a_3^2-(a_1^2+a_2^2)}]=\pm 1$ in this case, we have 
\begin{align}
\sum_{{\rm Im}E_p>0}\nu_p=-\sum_{{\rm Im}E_p<0}\nu_p=1.
\end{align}

\subsubsection{Case with $\sqrt{a_0^2+a_1^2+a_2^2}>a_3>\sqrt{a_1^2+a_2^2}$, or
$-\sqrt{a_1^2+a_2^2}>a_3>-\sqrt{a_0^2+a_1^2+a_2^2}$}
It holds that ${\rm sgn}[{\rm Im}E_\pm]={\rm sgn}[a_0]$ in these cases.  
Thus, ${\rm Im}E_+$ and ${\rm Im}E_-$ have the same sign, leading to
\begin{align}
\sum_{{\rm Im}E_p>0}\nu_p=-\sum_{{\rm Im}E_p<0}\nu_p=0.
\end{align}

\subsubsection{Case with $\sqrt{a_1^2+a_2^2}>a_3>-\sqrt{a_1^2+a_2^2}$ }
Since no gapless state exists in this case, we have
\begin{align}
\sum_{{\rm Im}E_p>0}\nu_p=-\sum_{{\rm Im}E_p<0}\nu_p=0.
\end{align}

\subsubsection{Case with $-\sqrt{a_0^2+a_1^2+a_2^2}>a_3$}
Because $a_3<-\sqrt{a_1^2+a_2}$, it holds that ${\rm sgn}[a_3\pm\sqrt{a_3^2-(a_1^2+a_2^2)}]=-1$.
We also have ${\rm sgn}\left[{\rm Im}E_{\pm}\right]={\rm sgn}[a_0\pm \sqrt{a_3^2-(a_1^2+a_2^2)}]=\pm 1$, which leads to
\begin{align}
\sum_{{\rm Im}E_p>0}\nu_p=-\sum_{{\rm Im}E_p<0}\nu_p=-1. 
\end{align}

In summary, we have
\begin{align}
\sum_{{\rm Im}E_p>0}\nu_p=-\sum_{{\rm Im}E_p<0}\nu_p=\left\{
\begin{array}{cl}
1, & \mbox{for $a_3>\sqrt{a_0^2+a_1^2+a_2^2}$},\\
0, & \mbox{for $\sqrt{a_0^2+a_1^2+a_2^2}>a_3>-\sqrt{a_0^2+a_1^2+a_2^2}$},\\
-1, & \mbox{for $-\sqrt{a_0^2+a_1^2+a_2^2}>a_3$},
\end{array}
\right. .
\end{align}

We compare this result with $n$.
The Hermitian matrix $iH\Gamma$ is given by
\begin{align}
i H\Gamma=-a_0\sigma_3+a_1\sigma_2-a_2\sigma_1-a_3,  
\end{align}
of which eigenvalues are
\begin{align}
{\cal E}_\pm=-a_3\pm\sqrt{a_0^2+a_1^2+a_2^2}. 
\end{align}
Thus, $n$ in Eq.(\ref{seq:npAIII0d}) is evaluated as 
\begin{align}
n=\left\{
\begin{array}{cl}
1, & \mbox{for $a_3>\sqrt{a_0^2+a_1^2+a_2^2}$},\\
0, & \mbox{for $\sqrt{a_0^2+a_1^2+a_2^2}>a_3>-\sqrt{a_0^2+a_1^2+a_2^2}$},\\
-1, & \mbox{for $-\sqrt{a_0^2+a_1^2+a_2^2}>a_3$},
\end{array}
\right. ,
\end{align}
which yields Eq.(\ref{eq:SThm1}) in Theorem.

\subsection{Class BDI$^{\dagger}$}
\label{sec:S3B}

Symmetry in class BDI$^\dagger$ is given by
\begin{align}
T H^T T^{-1}=H, \quad C H^* C^{-1}=- H, \quad TT^*=1, \quad CC^*=1, 
\end{align}
where $T$ and $C$ are unitary matrices with $CT^*=TC^*$, which define TRS$^\dagger$ and PHS$^\dagger$, respectively. 
By combining  TRS$^{\dagger}$ with PHS$^{\dagger}$, the Hamiltonian $H$ also has CS,
\begin{align}
\Gamma H^{\dagger}\Gamma^{-1}=-H, \quad \Gamma=TC^{*}, 
\end{align}  
from which the point gap topological number $n$ and the topological number $\nu_p$ for a gapless region are defined in the same manner as class AIII.
A general $2\times 2$ matrix in class BDI$^\dagger$ is given by
\begin{align}
H=i a_0\sigma_0+a_1 \sigma_1+i a_3 \sigma_3, \quad T=1, \quad C=\sigma_3,  
\label{seq:2x2BDIdagger}
\end{align}
where $a_\mu$ are real parameters. 
Theorem holds for $H$ in the above
since $H$ in Eq.(\ref{seq:2x2BDIdagger})
is a special case of $H$ in Eq(\ref{seq:2x2AIII}) with $a_2=0$.

\subsection{Class D$^{\dagger}$}
\label{sec:S3C}

The Hamiltonian $H$ in class D$^\dagger$ satisfies
\begin{align}
C H^* C^{-1}=- H, \quad CC^*=1, 
\label{seq:0dD}
\end{align}
with a unitary matrix $C$.
This relation implies that ${\rm det}(iH)$ is real, and
the point-gap $\mathbb{Z}_2$ invariant $n$ is defined by 
\begin{align}
(-1)^{n}={\rm sgn}[{\rm det}(iH)].
\end{align}
On the other hand, the presence or absence of a gapless state defines the
$\mathbb{Z}_2$ invariant $\nu$ for the gapless state, where the $\mathbb{Z}_2$ nature implies that the presence of an even number of gapless states trivializes $\nu$.

A general $2\times 2$ matrix in class D$^\dagger$ is given by
\begin{align}
H=ia_0\sigma_0+ia_1\sigma_1+a_2\sigma_2+ia_3\sigma_3, \quad C=1,
\end{align}
of which eigenvalues are 
\begin{align}
E_{\pm}=ia_0\pm \sqrt{a_2^2-a_1^2-a_3^2}.
\end{align}
When $a_1^2+a_3^2>a_2^2$, the eigenstates with eigenvalue $E_{\pm}$ become gapless in the sense of ${\rm Re}E_{\pm}=0$, and they have opposite signs in ${\rm Im}E_{\pm}$ if $a_1^2+a_3^2>a_0^2+a_2^2$.
Thus, taking into account the $\mathbb{Z}_2$ nature of the gapless states, we have
\begin{align}
\sum_{{\rm Im}E_p>0}\nu_p=-\sum_{{\rm Im}E_p<0}\nu_p=
\left\{
\begin{array}{cl}
0 \quad {\rm mod}.2, & \mbox{for $a_1^2+a_3^2>a_0^2+a_2^2$},\\
1 \quad {\rm mod}.2, &  \mbox{for $a_1^2+a_3^2<a_0^2+a_2^2$}.
\end{array}
\right.
\end{align} 
We also find
\begin{align}
{\rm det}(iH)=a_0^2+a_3^2-a_1^2-a_3^2, 
\end{align}
from which $n$ is evaluated as
\begin{align}
(-1)^{n}=\left\{
\begin{array}{cl}
1, & \mbox{for $a_1^2+a_3^2>a_0^2+a_2^2$,}\\
-1, & \mbox{for $a_1^2+a_3^2<a_0^2+a_2^2$}.
\end{array}
\right.
\end{align}
Thus, Theorem holds.

\subsection{Class DIII$^{\dagger}$}
\label{sec:S3D}

The Hamiltonian $H$ in class DIII$^\dagger$ satisfies
\begin{align}
T H^T T^{-1}=H, \quad C H^* C^{-1}=- H, \quad TT^*=-1, \quad CC^*=1, 
\end{align}
where $T$ and $C$ are unitary matrices with $CT^*=TC^*$.
Because of TRS$^\dagger$, the matrix $HT$ is antisymmetric, so the Pfaffian ${\rm Pf}(HT)$ can be defined.
For a $2p\times 2p$ antisymmetric matrix $HT$, we can also show
\begin{align}
[{\rm Pf}(HT)]^*=(-1)^p{\rm det}(C^*){\rm Pf}(HT).
\end{align} 
With an appropriate basis satisfying ${\rm det}(C^*)=(-1)^p$, 
the point-gap $\mathbb{Z}_2$ invariant $n$ is defined as 
\begin{align}
(-1)^{n}={\rm sgn}[{\rm Pf}(HT)].
\end{align}
Like the class D$^\dagger$ case, the presence or absence of a gapless state defines the $\mathbb{Z}_2$ invariant for the gapless state.
An important difference is the Kramers degeneracy.
Because each eigenstate of $H$ doubly degenerates due to the Kramers Theorem for TRS$^\dagger$, an even number of the Kramers pairs trivializes $\nu$.

To obtain a non-trivial example, we need a $4\times 4$ matrix. A general $4\times 4$ matrix in class DIII$^\dagger$ is
\begin{align}
H=ia_{00} \tau_0\sigma_0+ia_{10}\tau_1\sigma_0+ia_{30}\tau_3\sigma_0+a_{21}\tau_2\sigma_1
+a_{22}\tau_2\sigma_2+ia_{23}\tau_2\sigma_3, \quad T=\tau_0\sigma_2, \quad C=\tau_0\sigma_1, 
\end{align}
The eigenvalues with Kramers degeneracy are
\begin{align}
E_{\pm}=ia_{00}\pm \sqrt{a_{21}^2+a_{22}^2-a_{10}^2-a_{30}^2-a_{23}^2}, 
\end{align}
which becomes gapless in the sense of ${\rm Re}E_{\pm}=0$ for $a_{10}^2+a_{30}^2+a_{23}^2>a_{21}^2+a_{22}^2$.
These gapless states have opposite signs in ${\rm Im}E_{\pm}$ when $a_{10}^2+a_{30}^2+a_{23}^2>a_{00}^2+a_{21}^2+a_{22}^2$, leading to 
\begin{align}
\sum_{{\rm Im}E_p>0}\nu_p=-\sum_{{\rm Im}E_p<0}\nu_p=
\left\{
\begin{array}{cl}
0 \quad \mbox{mod. 2}, & \mbox{for $a_{10}^2+a_{30}^2+a_{23}^2<a_{00}^2+a_{21}^2+a_{22}^2$},\\
1 \quad \mbox{mod. 2}, & \mbox{for $a_{10}^2+a_{30}^2+a_{23}^2>a_{00}^2+a_{21}^2+a_{22}^2$}.
\end{array}
\right. 
\end{align} 
We also find 
\begin{align}
{\rm Pf}(HT)=a_{00}^2+a_{21}^2+a_{22}^2-a_{30}^2-a_{10}^2-a_{23}^2, 
\end{align}
so we obtain
\begin{align}
(-1)^{n}=
\left\{
\begin{array}{cl}
1, & \mbox{for $a_{10}^2+a_{30}^2+a_{23}^2<a_{00}^2+a_{21}^2+a_{22}^2$},\\
-1, &\mbox{for $a_{10}^2+a_{30}^2+a_{23}^2>a_{00}^2+a_{21}^2+a_{22}^2$}.
\end{array}
\right. 
\end{align}
Thus, Theorem holds.

\subsection{Class CII$^{\dagger}$}
\label{sec:S3E}

The Hamiltonian in class CII$^\dagger$ obeys 
\begin{align}
T H^T T^{-1}=H, \quad C H^* C^{-1}=- H, \quad TT^*=-1, \quad CC^*=-1, 
\end{align}
where $T$ and $C$ are unitary matrices with $CT^*=TC^*$. Combining TRS$^\dagger$ with PHS$^\dagger$, we also have CS,
\begin{align}
\Gamma H^{\dagger} \Gamma^{-1}=-H, \quad \Gamma=TC^*, 
\end{align}
which yields that $iH\Gamma$ is Hermitian. The matrix $iH\Gamma$ also has its own ordinary TRS defined by $C$
\begin{align}
C[iH\Gamma]^*C^{-1}=iH\Gamma,
\end{align}
which results in the Kramers degeneracy for eigenstates of $iH\Gamma$.
The point gap topological invariant $n$ is defined by
\begin{align}
n=\frac{1}{2}\left[N_-(iH\Gamma)-N_+(iH\Gamma)\right], 
\end{align}
where $N_{-}(iH\Gamma)$ ($N_+(iH\Gamma)$) is the number of Kramers pairs of negative (positive) eigenstates in $iH\Gamma$.
On the other hand, the topological invariant $\nu_p$ for a gapless state is defined like class AIII, while the  Kramers degeneracy in $H$ needs to be considered.
For right and left eigenstates of $H$,
\begin{align}
H|u_{2p-1}\rangle=E_{2p-1}|u_{2p-1}\rangle, \quad &\langle\!\langle u_{2p-1}|H=E_{2p-1} \langle\!\langle u_{2p-1}|,
\\
H|u_{2p}\rangle=E_{2p}|u_{2p}\rangle, \quad  &\langle\!\langle u_{2p}|H=E_{2p} \langle\!\langle u_{2p}|,
\end{align}
we have the Kramers degeneracy,
\begin{align}
E_{2p-1}=E_{2p}, \quad |u_{2p}\rangle=T|u_{2p-1}^*\rangle\!\rangle. 
\end{align}
When the Kramers pair is gapless in the sense of ${\rm Re}E_{2p-1}={\rm Re}E_{2p}=0$, they have a common non-Hermitian chirality defined by 
\begin{align}
\Gamma|u_{2p-1}\rangle\!\rangle=\pm |u_{2p-1}\rangle, \quad \Gamma|u_{2p}\rangle\!\rangle=\pm |u_{2p}\rangle.
\end{align}
The common chirality defines the topological invariant $\nu_p$ for the Kramers pair of gapless states.

A general $4\times 4$ matrix in class CII$^{\dagger}$ is 
\begin{align}
H=ia_{00}\tau_0\sigma_0+a_{10}\tau_1\sigma_0+ia_{30}\tau_3\sigma_0+a_{21}\tau_2\sigma_1
+a_{22}\tau_2\sigma_2+a_{23}\tau_2\sigma_3, \quad T=\tau_0\sigma_2, \quad C=\tau_3\sigma_2, 
\end{align}
where $T$ and $C$ are unitary matrices with $CT^*=TC^*$. The eigenvalues of $H$ are
\begin{align}
E_{\pm}=ia_{00}\pm\sqrt{a_{10}^2+a_{21}^2+a_{22}^2+a_{23}^2-a_{30}^2}, 
\end{align}
with the Kramers degeneracy. 
When $a_{30}^2>a_{10}^2+a_{21}^2+a_{22}^2+a_{23}^2$, 
they are gapless in the sense of ${\rm Re}E_{\pm}=0$, each Kramers pair satisfies  
\begin{align}
\Gamma|\pm\rangle\!\rangle=\pm {\rm sgn}\left[a_{30}\pm \sqrt{a_{30}^2-a_{10}^2-a_{21}^2-a_{22}^2-a_{23}^2}
\right]|\pm\rangle. 
\end{align}
Furthermore, when $a_{30}^2>a_{00}^2+a_{10}^2+a_{21}^2+a_{22}^2+a_{23}^2$, 
they have different signs in ${\rm Im}E_{\pm}$. 
Thus, we have
\begin{align}
\sum_{{\rm Im}E_{2p}>0}\nu_p= -\sum_{{\rm Im}E_{2p}<0}\nu_p=
\left\{
\begin{array}{cl}
1, & \mbox{for $a_{30}^2>a_{00}^2+a_{10}^2+a_{21}^2+a_{22}^2+a_{23}^2$}, \\
0, & \mbox{for $a_{30}^2<a_{00}^2+a_{10}^2+a_{21}^2+a_{22}^2+a_{23}^2$}.
\end{array}
\right. 
\end{align}
On the other hand, the Hermitian matrix $iH\Gamma$ is
\begin{align}
iH\Gamma=-a_{00}\tau_3\sigma_0+a_{10}\tau_2\sigma_0-a_{30}\tau_0\sigma_0
-a_{21}\tau_1\sigma_1-a_{22}\tau_1\sigma_2-a_{23}\tau_1\sigma_3, 
\end{align}
of which eigenvalues are
\begin{align}
{\cal E}_{\pm}=-a_{30}\pm\sqrt{a_{00}^2+a_{10}^2+a_{21}^2+a_{22}^2+a_{23}^2}. 
\end{align}
Thus, $n$ of this model is
\begin{align}
n=\left\{
\begin{array}{cl}
1, & \mbox{for $a_{30}^2>a_{00}^2+a_{10}^2+a_{21}^2+a_{22}^2+a_{23}^2$}, \\
0, & \mbox{for $a_{30}^2<a_{00}^2+a_{10}^2+a_{21}^2+a_{22}^2+a_{23}^2$}.
\end{array}
\right. 
\end{align}
Therefore, Theorem holds.

\section{Extended Nielsen-Ninomiya Theorem in periodically driven systems}
\label{sec:SF}

Using the duality relation
\begin{align}
H({\bm k})=i U_{\rm F}({\bm k}),    
\end{align}
we obtain the extended Nielsen-Ninomiya Theorem for periodically driven systems.

\vspace{3ex}
\noindent
{\bf Theorem' } 
Let $H_{\rm F}({\bm k})$ be a $d$-dimensional Floquet Hamiltonian in an AZ class.
Then, gapless points in $H_{\rm F}({\bm k})$ obey the following relation,
\begin{align}
&n=\sum_{\alpha}\nu^0_{\alpha}
=-\sum_{\alpha}\nu^{\pi}_{\alpha},  &\mbox{for even $d$}, 
\label{eq:SThm1'even}
\\
&n=\sum_{\alpha}\nu^0_{\alpha}
=\sum_{\alpha}\nu^{\pi}_{\alpha},  &\mbox{for odd $d$}, 
\label{eq:SThm1'odd}
\end{align}
where $n$ is the bulk topological invariant of  $U_{\rm F}=e^{-iH_{\rm F}({\bm k})\tau}$.
In classes A, AI, and AII,  $\nu^{0 (\pi)}_\alpha$ is  the topological charge defined on the $\alpha$-th Fermi surface at the quasi energy $\epsilon=0 (\pi)$.
In other AZ classes, $\nu^{0(\pi)}_\alpha$ is the topological charge of the $\alpha$-th gapless point at $\epsilon=0 (\pi)$.

\vspace{3ex}

In classes A, AI, and AII, we can change the Fermi energy arbitrary by multiplying $U_{\rm F}({\bm k})$ by the phase factor $e^{i\mu\tau}$.
We can also neglect the minus sign in the last term in Eq.(\ref{eq:SThm1'even}) in these AZ classes
because
$n$ and $\nu_\alpha^{0 (\pi)}$ in those AZ classes are $\mathbb{Z}_2$ numbers in even dimensions. Thus, the following Corollary holds.

\vspace{3ex}
\noindent
{\bf Corollary} 
For classes A, AI, and AII, we have 
\begin{align}\label{eq:SCorollary}
n=\sum_{\alpha}\nu^\mu_{\alpha},
\end{align}
where $\nu^{\mu}_\alpha$ is  the topological charge defined on the $\alpha$-th Fermi surface at the quasi energy $\epsilon=\mu$.

\vspace{3ex}
Here we have omitted the term corresponding to the last terms in Eqs.(\ref{eq:SThm1'even}) and (\ref{eq:SThm1'odd}) since it is just a particular case of Eq.(\ref{eq:SCorollary}).
For class A in three dimensions, Corollary implies
\begin{align}
w_3=\sum_{\alpha}{\rm Ch}^\mu_\alpha,    
\end{align}
where ${\rm Ch}_\alpha^\mu$ is the Chern number of the $\alpha$-th Fermi surface at the quasi energy $\mu=\epsilon$.

\section{NIELSEN-NINOMIYA THEOREM}
\label{sec:S3}

The original Nielsen-Ninomiya (NN) theorem \cite{Nielsen81,Nielsen81ii} states that the total chirality of Weyl points in a Hermitian Hamiltonian should be zero.
Here we reformulate the theorem differently, which is more convenient to describe gapless modes in non-Hermitian and periodically driven systems.

Let us consider a Hermitian Hamiltonian $H({\bm k})$ with eigenvalues $E_p({\bm k})$ in three dimensions.
For this Hamiltonian, the following relation holds,
\begin{align}
\sum_{p\alpha}{\rm Ch}(S_{p\alpha})=0,
\label{Seq:NN}
\end{align}
where $S_{p\alpha}$ is the Fermi surface defined by $\{{\bm k}\in S_{p\alpha}|E_p({\bm k})=0\}$ and ${\rm Ch}(S_{p\alpha})$ is the Chern number on $S_{p\alpha}$.

\vspace{1ex}
\noindent

\begin{figure}[htbp]
\centering
\includegraphics[width=50mm]{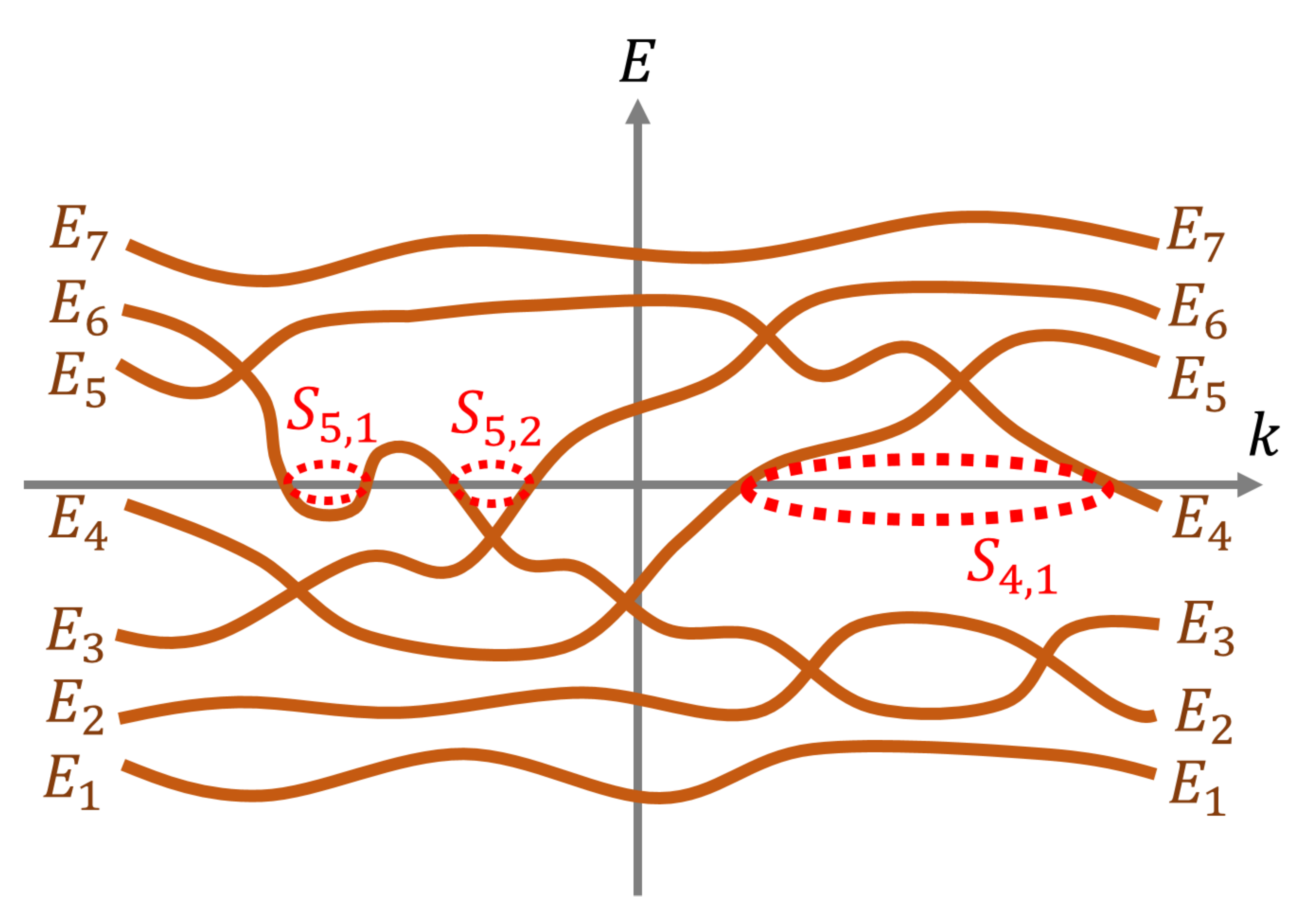} 
\caption{Typical band dispersion in Hermitian systems. In this case, Eq.~(\ref{Seq:NN}) states 
${\rm Ch}(S_{5,1})+{\rm Ch}(S_{5,2})+{\rm Ch}(S_{4,1})=0$.
}
	\label{fig:NN3}
\end{figure}

Proof: 
First, we order $E_p({\bm k})$ as $E_1({\bm k})\le E_2({\bm k}) \le E_3({\bm k}) \dots$ as illustrated in Fig.~\ref{fig:NN3}.
Then, we continuously deform each energy $E_p({\bm k})$ so as to satisfy either $E_p({\bm k})>0$ or $E_p({\bm k})<0$ in the whole Brillouin zone. After the deformation, Eq.~(\ref{Seq:NN}) obviously holds because there is no Fermi surface and there is no term on the left-hand side.
Therefore, if the left-hand side of Eq.~(\ref{Seq:NN}) is invariant during the above deformation, Eq.~(\ref{Seq:NN}) holds. We show this is indeed the case by moving bands upward one by one.

\begin{figure}[hbtp]
\centering
\includegraphics[width=45mm]{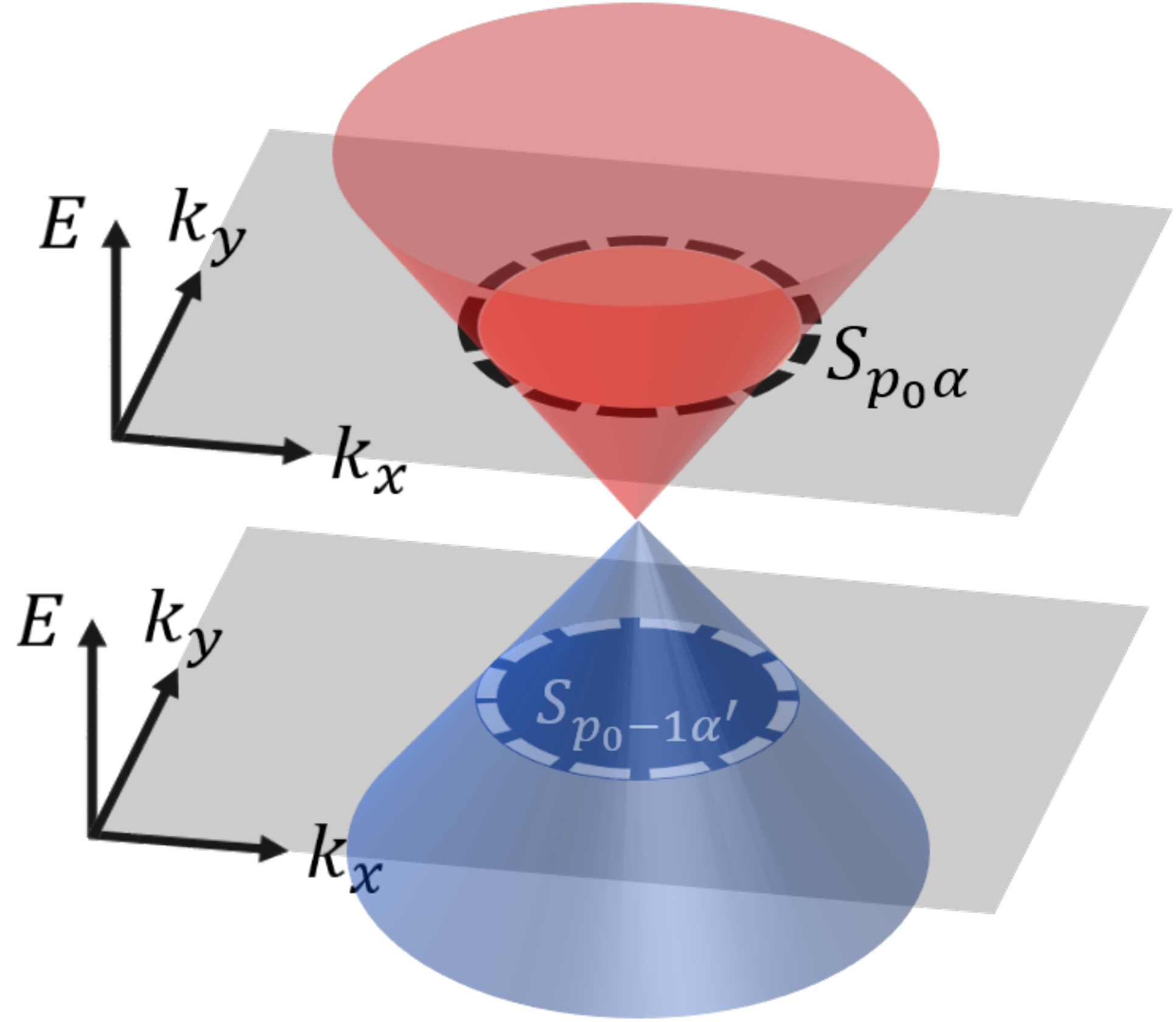} 
\caption{Energy dispersion near a degenerate point (Weyl point). We omit the $k_z$ dependence for simplicity. 
The upper (lower) gray plane indicates the Fermi energy $E=0$ before (after) a band with the Weyl point moves upward. 
When the band moves upward, the Fermi surface $S_{p_0\alpha}$ shrinks to the Weyl point, then a new Fermi surface $S_{p_0-1\alpha'}$ is created in a lower band $p_0-1$.
Note that the orientation of $S_{p_0-1\alpha'}$ is opposite to that of $S_{p_0\alpha}$ because of the difference in their Fermi velocities.
}
	\label{fig: NHWeylERing}
\end{figure}

Let us consider a metallic band that hosts at least one Fermi surface at the Fermi energy $E=0$.  
When we move the band upward, the following four processes may happen. 
(a) A Fermi surface shrinks and vanishes smoothly.
(b) A Fermi surface merges into another Fermi surface or splits into two Fermi surfaces.
(c) A new Fermi surface is created smoothly.
(d) A Fermi surface shrinks to a Weyl point then moves to a lower band. 
During the first three processes, the left-hand side of Eq.~(\ref{Seq:NN}) is invariant because the Chern number cannot change during such smooth deformations.
Notably, the last process also keeps the left-hand side of Eq.~(\ref{Seq:NN}) invariant because we have
\begin{align}
{\rm Ch}(S_{p_0\alpha})={\rm Ch}(S_{p_0-1\alpha'}),
\end{align}
where $S_{p_0\alpha}$ is the Fermi surface shrinking into the Weyl point and 
$S_{p_0-1\alpha'}$ is the Fermi surface created on the lower band in this process.
(This equation is directly shown by the Hamiltonian $H(\bm{k})=\sum_{ij} a_{ij}k_i \sigma_j$ describing the Weyl point.)
Therefore, the left-hand side of Eq.~(\ref{Seq:NN}) is invariant when we move all metallic bands upward above the Fermi energy. Consequently, we have Eq.~(\ref{Seq:NN}).

\section{ NON-HERMITIAN WEYL Semimetal}
\label{sec:S4}

\subsection{Extended Nielsen-Ninomiya theorem}
Consider the non-Hermitian Weyl semimetal given by \begin{align}
\label{eq:BesWeylS}
H({\bm k})=\left(d_0+{\bm d}({\bm k})\cdot{\bm \sigma}\right) \tau_{1} 
+\left(m({\bm k})+i\gamma\right) \tau_{3}   -i \gamma_0\tau_{0},
\end{align}
with 
\begin{align}
d_i({\bm k})=\sin k_i, \quad
m({\bm k})=m_0+\cos k_1+\cos k_2 +\cos k_3,
\end{align}
where $d_0$, $m_0$, $\gamma$, and $\gamma_0$ are real constants. The band energies of this model are
\begin{align}
&E_1({\bm k})=\sqrt{(|{\bm d}({\bm k})|+d_0)^2+(m({\bm k})+i\gamma)^2}
-i\gamma_0, 
\quad
E_2({\bm k})=\sqrt{(|{\bm d}({\bm k})|-d_0)^2+(m({\bm k})+i\gamma)^2}
-i\gamma_0,\nonumber\\
&E_3({\bm k})=-\sqrt{(|{\bm d}({\bm k})|+d_0)^2+(m({\bm k})+i\gamma)^2}
-i\gamma_0,\quad
E_4({\bm k})=-\sqrt{(|{\bm d}({\bm k})|-d_0)^2+(m({\bm k})+i\gamma)^2}
-i\gamma_0.
\label{eq:Sband}
\end{align}

\begin{figure}[htbp]
\centering
\includegraphics[width=160mm]{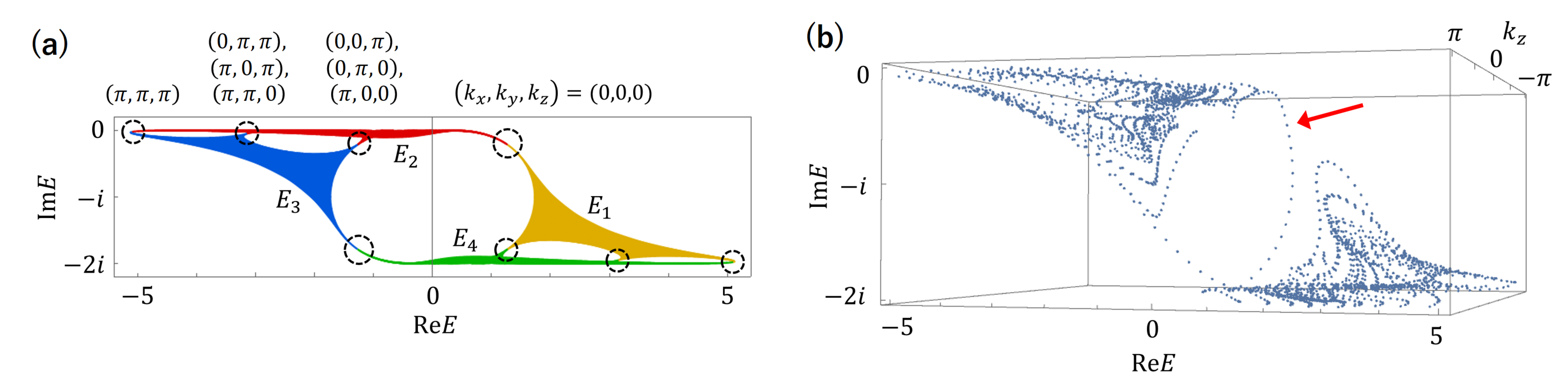} 
\caption{(a) Complex energy spectrum of the non-Hermitian Weyl semimetal in Eq.~(\ref{eq:BesWeylS}) with $d_0=\gamma=\gamma_0=1$ and $m_0=-2$. 
Different colors distinguish different bands in Eq.~(\ref{eq:Sband}), and the red circles emphasize  Weyl points.
(b) Complex energy spectrum of Eq.~(\ref{eq:BesWeylS}) under a magnetic field $B_z$. The red arrow indicates a right-going mode originating from the Weyl point at ${\bm k}=(0,0,0)$. 
The system size is $L_x=L_y=L_z=30$ and the magnetic flux is $B_z/2\pi=1/10$ in (b).}
	\label{fig:BesWeylEnergyS}
\end{figure}

Figure ~\ref{fig:BesWeylEnergyS} illustrates the energy spectrum having Weyl points at ${\bm k}=(0,0,0)$, $(\pi, 0, 0)$, $(0, \pi, 0)$, $(0,0,\pi)$, $(\pi, \pi, 0)$, $(\pi, 0, \pi)$, $(0, \pi, \pi)$, $(\pi,\pi,\pi)$. 
The spectrum has a point gap at $E =-i\gamma_0\equiv E_{\rm F}$.
Let us evaluate the topological charge of the Weyl points inside the Fermi surfaces defined by ${\bm k}$ satisfying ${\rm Re}(E({\bm k})-E_{\rm F})=0$.
When $d_0=\gamma=\gamma_0=1$ and $m_0=-2$,  only $E_2({\bm k})$ and $E_4({\bm k})$ bands host such Fermi surfaces, which we denote by $S_2$ and $S_4$, respectively.  The Fermi surface $S_2$ ($S_4$) has an imaginary part of the energy higher (lower) than the imaginary part of $E_{\rm F}$.

The right eigenfunction of $H({\bm k})$ with the eigenenergy $E_2({\bm k})$ is given by
\begin{align}
|\psi_2({\bm k})\rangle=
\frac{1}{\sqrt{2|{\bm d}({\bm k})|(|{\bm d}({\bm k})|-d_3({\bm k}))}}
\left(
\begin{array}{c}
d_3({\bm k})-|{\bm d}({\bm k})|\\
d_1({\bm k})+i d_2({\bm k})
\end{array}
\right)_\sigma
\otimes
\left(
\begin{array}{c}
m({\bm k})+i\gamma+E_2({\bm k})\\
d_0-|{\bm d}({\bm k})|
\end{array}
\right)_\tau.
\end{align}
We also have a similar expression for the corresponding left eigenfunction $\langle\!\langle \psi_2({\bm k})|$,  which is normalized as $\langle\!\langle \psi_2({\bm k})|\psi_2({\bm k})\rangle=1$. 
The Chern number of the Fermi surface $S_2$ is given by
\begin{align}
{\rm Ch}(S_2)=\frac{1}{2\pi i}\int_{S_2}  (\nabla \times \bm{A}(\bm{k}) )\cdot \text{d}{\bm S},
\end{align}
where ${\bm A}({\bm k})=\langle\!\langle \psi_2({\bm k})|\nabla \psi_2({\bm k})\rangle$ and the area element $d{\bm S}$ points to the direction of the Fermi velocity $\nabla {\rm Re}E_2({\bm k})|_{{\bm k}\in S_2}$.
As $S_2$ encloses a Weyl point at ${\bm k}=(0,0,0)$ in the upper right side of Fig.~\ref{fig:BesWeylEnergyS},  we find that ${\rm Ch}(S_2)=1$.
In a similar manner, we also obtain ${\rm Ch}(S_4)=-1$.
On the other hand, we numerically evaluate the 3D winding number $w_3$ given by  
\begin{align}
w_3=-\frac{1}{24\pi^2}\int_{\rm BZ}{\rm tr}
\left[(H-E_{\rm F})^{-1}d(H-E_{\rm F})\right]^3,    
\end{align}
and find that $w_3=1$ for $d_0=\gamma=\gamma_0=1$ and $m_0=-2$.
Therefore, Theorem 2 in the main text holds in this model.

\subsection{Weyl point under a magnetic field}

\begin{figure}[htbp]
 \begin{center}
  \includegraphics[width=120mm]{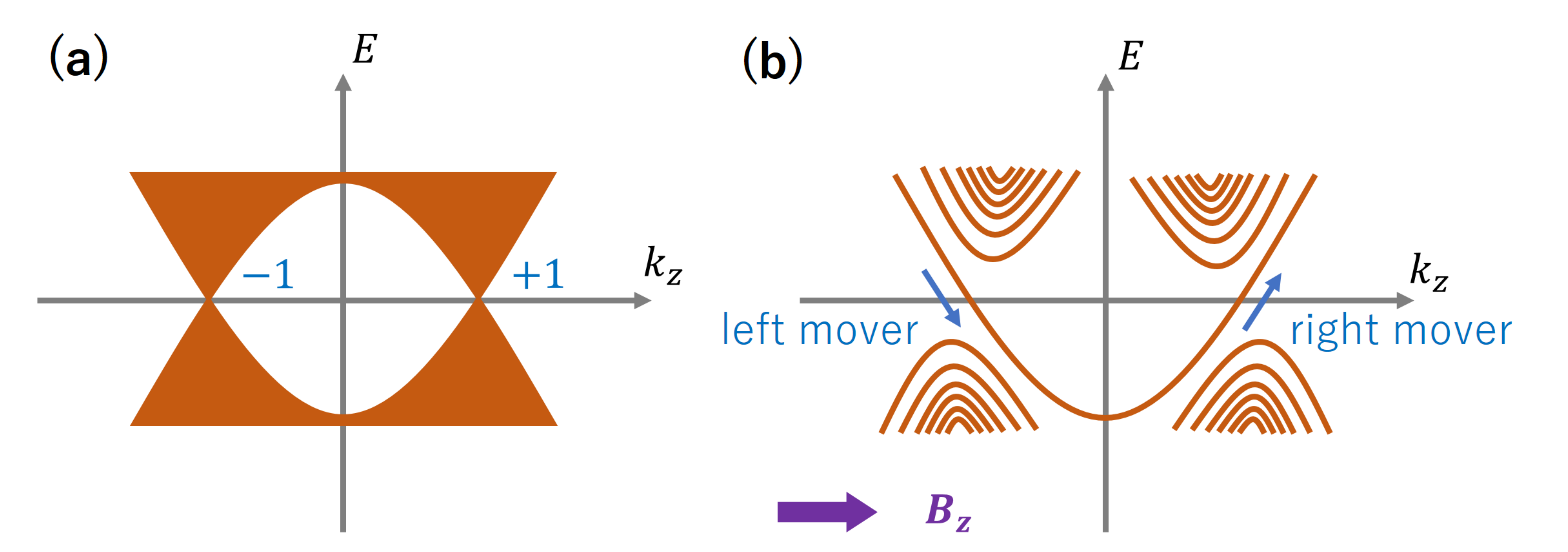}
 \end{center}
 \caption{Weyl points (a) without and (b) with a magnetic field. (a) Weyl points with $\pm 1$ chiralities. (b) The Weyl point with chirality 1 ($-1$) becomes a right (left) moving mode with $(B_z/2\pi) L_x L_y$-fold degeneracy. }
 \label{fig:WeylBz}
\end{figure}

We review  a Weyl point in an applied magnetic field. Consider the Weyl Hamiltonian with $+1$ chirality,
\begin{align}\label{eq:Weyl}
H=k_x \sigma_x +k_y \sigma_y +k_z \sigma_z.
\end{align}
Under a magnetic field $B_z$, {\it i.e.,} the vector potential $\bm{A}=(0,B_z x,0)$, the Hamiltonian reads
\begin{align}\label{eq:WeylB}
\hat{H}=-i\partial_x \sigma_x +(k_y-eB_z x) \sigma_y +k_z \sigma_z=
\pmat{k_z & -i\partial_x - i (k_y-eB_z x) \\ -i\partial_x + i (k_y-eB_z x) & -k_z},
\end{align}
where $e$ is the electric charge of the Weyl fermion.
For $eB_z>0$, $\hat{H}$ is recast into
\begin{align}
\hat{H} =
\pmat{k_z & \sqrt{2 e B_z}\hat{a}^\dag
 \\ \sqrt{2e B_z}\hat{a} & -k_z},
\label{eq:Weyla}
\end{align}
in terms of the annihilation and creation operators,
\begin{align}
\hat{a}=\frac{-i\partial_x+i(k_y-eB_z x)}{\sqrt{2e B_z}}, \quad
\hat{a}^\dag=\frac{-i\partial_x-i(k_y-eB_z x)}{\sqrt{2e B_z}}, \quad [\hat{a},\hat{a}^\dag]=1.
\end{align}
%
The Hamiltonian $\hat{H}$ has a single right moving chiral mode with $E=k_z$ and non-chiral gapped modes with $E=\pm \sqrt{k_z^2+2e B_z p} \ (p=1,2,3,\ldots)$, as illustrated in Fig.\ref{fig:WeylBz}.
For the right-moving mode with $E=k_z$, the eigen equation $\hat{H}|\psi\rangle=E|\psi\rangle$ with $|\psi\rangle=(|\psi_1\rangle,|\psi_2\rangle)$ leads to
\begin{align}
\hat{a}\ket{\psi_1}=0, \quad 
\ket{\psi_2}=0, 
\end{align}
from which  we have the wave function of the right moving mode,
\begin{align}
\psi_1(x)=\left(\frac{e B_z}{\pi}\right)^{1/4} \exp\left[-\frac{eB_z}{2}\left(x-\frac{k_y}{eB_z}\right)^2\right],
\quad \psi_2(x)=0.
\end{align}
The wave function $\psi_1(x)$ has the center at $x_c=k_y/eB_z$.
In the periodic boundary condition in the $x$ and $y$ directions, $x_c$ and $k_y$ satisfy $0< x_c\le L_x$ and $k_y=2\pi n_y/L_y$, respectively, of which compatibility leads to $n_y=1, \dots, (eB_z/2\pi)L_x L_y$.
Therefore, the right-moving mode has $(eB_z/2\pi) L_x L_y$-fold degeneracy. 
Note that $(eB_z/2\pi)L_xL_y$ is an integer in order for the periodic boundary conditions to be consistent with
magnetic translation symmetry.
In a similar manner, for $eB_z<0$, we have a left-moving mode $E=-k_z$ with $-(eB_z/2\pi)L_x L_y$ degeneracy.

For the Weyl Hamiltonian with $-1$ chirality, we have 
a left-moving mode $E=-k_z$ with $(eB_z/2\pi) L_x L_y$-fold degeneracy when $eB_z>0$, and a right-moving mode $E=k_z$ with $-(eB_z/2\pi) L_x L_y$-fold degeneracy when $eB_z<0$.

\subsection{Chiral magnetic skin effect}
\label{sec:S:CMSE}

\begin{figure}[tbhp]
 \begin{center}
  \includegraphics[width=180mm]{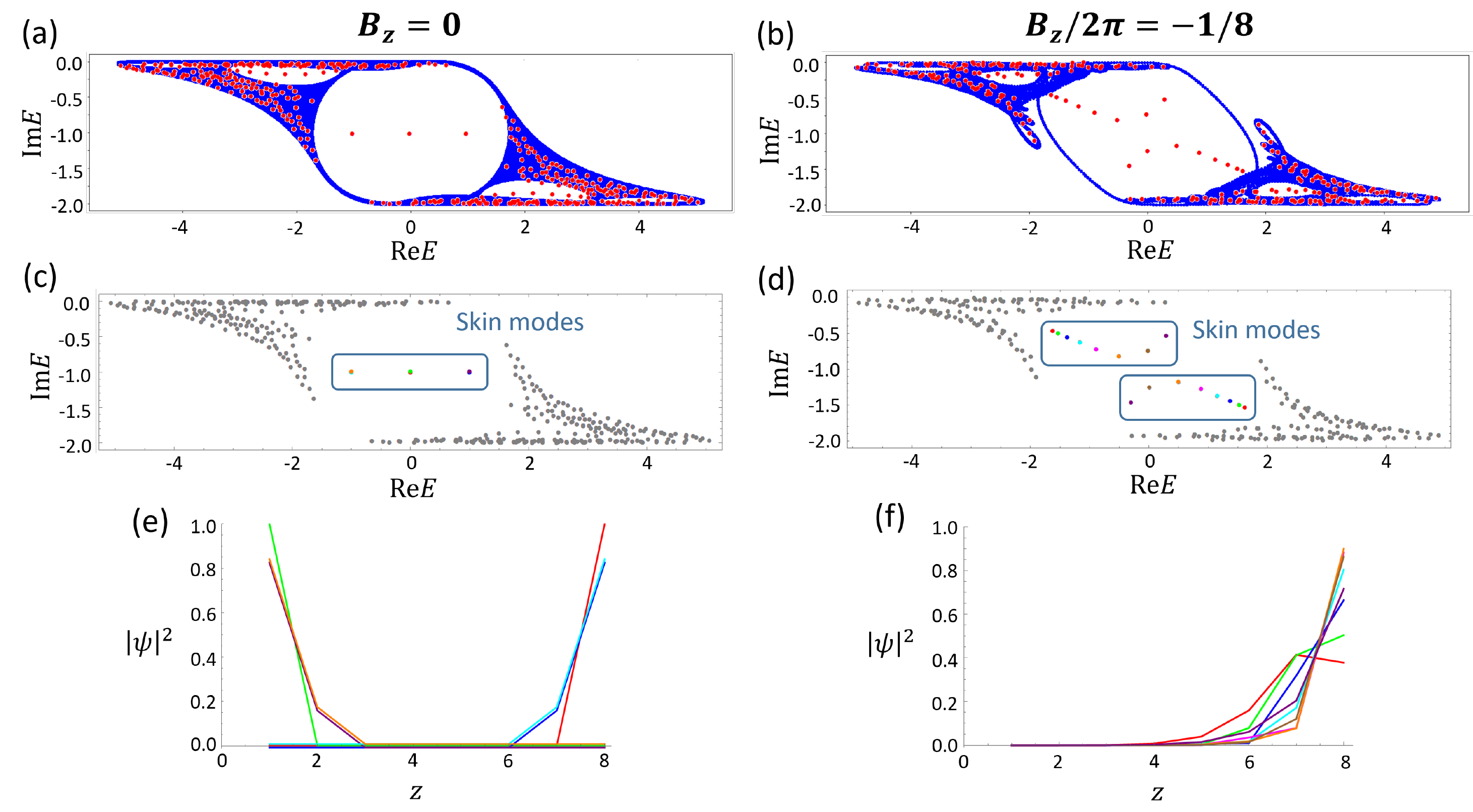}
\end{center}
 \caption{Chiral magnetic skin effect in the non-Hermitian Weyl semimetal of  Eq.(\ref{eq:BesWeylS}). (a,b) Energy spectra of Eq.(\ref{eq:BesWeylS}) with $d_0=\gamma=\gamma_0=1$, $m_0=-2$ (blue) in full periodic boundary conditions and (red) in the open boundary condition in the $z$ direction (zOBC). (c,d,e,f) Skin modes in zOBC with (c,e) $B_z=0$ and (d,f) $B_z/2\pi=-1/8$. 
 Colors in (c) and (d) correspond to the same colors in (e) and (f).
 The zOBC spectra and wave functions are calculated with the system size $L_x=L_y=L_z=8$. 
 All the skin modes in (f) under the magnetic field are localized at the right edge, as expected from the chiral magnetic effect.
 }
 \label{fig:CMSE}
\end{figure}

This section compares the complex energy spectra of the non-Hermitian Weyl semimetal in Eq.(\ref{eq:BesWeylS}) with and without a magnetic field and shows the chiral magnetic skin effect.

We first show the energy spectra of Eq.(\ref{eq:BesWeylS}) without a magnetic field in Fig.~\ref{fig:CMSE} (a). The blue region is the energy spectrum with the periodic boundary conditions in all directions, and the red dots are that with the open boundary condition in the $z$-direction and the periodic boundary conditions in other directions. Skin modes with  $E=-i$ and $E=\pm 1-i$ appear under the open boundary condition. 
This skin effect is nothing to do with the chiral magnetic skin effect. Instead, it originates from the $\mathbb{Z}_2$ number of the model Hamiltonian in Eq.~(\ref{eq:BesWeylS}).
Because of the reciprocity (transpose version of time-reversal symmetry) of $H({\bm k})$ in Eq.~(\ref{eq:BesWeyl}), $R H^T({\bm k}) R^{-1}=H(-{\bm k})$, $R=i\sigma_2\tau_0$, we can define the $\mathbb{Z}_2$ number for $H(0,0,k_z)$, 
\begin{eqnarray}
(-1)^{\nu(E)}={\rm sgn}\left\{\frac{{\rm Pf}[(H(0,0,\pi)-E)R]}{{\rm Pf}[(H(0,0,0)-E)R]}\exp\left[-\frac{1}{2}\int_{k_z=0}^{k_z=\pi}d\log\det[(H(0,0,k_z)-E)R] \right]
\right\}, 
\end{eqnarray} 
which is non-trivial, $\nu(E)=1$, when $E$ is in the center area 
enclosed by the blue spectrum in Fig.\ref{fig:CMSE} (a).
Thus, as a one-dimensional system in the $z$-direction, $H(0,0,k_z)$ exhibits 
the $\mathbb{Z}_2$ skin effect protected by the reciprocity \cite{OKSS20}.
By directly solving the (right) Schr\"{o}dinger equation, we obtain the skin modes with  $E=-i$ as
\begin{align}
\ket{\downarrow}_\sigma \otimes \ket{\leftarrow}_\tau\otimes\ket{z=0},
\quad
\ket{\uparrow}_\sigma \otimes \ket{\leftarrow}_\tau\otimes \ket{z=L_z},
\end{align}
each of which has two-fold degeneracy, and those with $E=\pm 1-i$ as 
\begin{align}
&    \mp i\ket{\downarrow}_\sigma \otimes\ket{\rightarrow}_\tau\otimes\ket{z=0} + \ket{\downarrow}_\sigma\otimes \ket{\leftarrow}_\tau \otimes(2 \ket{z=0} -i\ket{z=1}),
\nonumber\\
&\mp i\ket{\uparrow}_\sigma \otimes\ket{\rightarrow}_\tau\otimes\ket{z=L_z} + \ket{\uparrow}_\sigma\otimes \ket{\leftarrow}_\tau \otimes(2 \ket{z=L_z} -i\ket{z=L_z-1}),
\end{align}
each of which has $(L_z-1)$-fold degeneracy.  Here $\ket{\uparrow}_\sigma$ ($\ket{\downarrow}_\sigma$) and $\ket{\rightarrow}_\tau$ ($\ket{\leftarrow}_\tau$) are the $\sigma_z=1$ ($\sigma_z=-1$) eigenstate and the $\tau_y=1$ ($\tau_y=-1$) eigenstate, respectively, and $\ket{z=i}$ is the wave function localized at the $i$-th site in the $z$-direction. 
The skin modes localized at $z=0, L_z$ form Kramers pairs of the reciprocity $R$, and when one applies a magnetic field, they are mixed and disappear into the bulk.

In the presence of a magnetic field, however, different skin modes appear. 
In Fig.~\ref{fig:CMSE} (b), we show the energy spectra of Eq.~(\ref{eq:BesWeylS}) with a magnetic field in the $z$-direction.  
This data clearly show skin modes.
In contrast to the $\mathbb{Z}_2$ skin effect in the above, 
all the skin modes, in this case, are localized at $z=L_z$, as shown in Fig.~\ref{fig:CMSE} (f).
The present skin effect originates from the one-dimensional winding number $w_1$ \cite{OKSS20}.
As discussed in the main text, a non-zero $w_3$ induces a non-zero $w_1$ in the form of Eq. (\ref{eq:relation}) under a weak magnetic field, 
which provides the skin effect.
The existence of the skin modes is also consistent with the chiral magnetic effect since
the unidirectional movement of wave packets in Figs.\ref{fig:BesWeylEnergy} (c) and (d) results in an accumulation of bulk states at $z=L_z$ in the presence of the boundary.

We have also examined and confirmed the chiral magnetic skin effect in a different model.  More details of the chiral magnetic skin effect will be reported elsewhere \cite{Nakamura21}.

\section{2D CLASS AIII periodically driven system}
\label{sec:S5}

In this section, we consider a 2D periodically driven system in class AIII. 
Chiral symmetry $\Gamma H_{\rm F} \Gamma^{-1}=-H_{\rm F}$ in class AIII implies
$\Gamma U_{\rm F}^\dag \Gamma^{-1}=U_{\rm F}$, and thus $\Gamma U_{\rm F}$ is Hermitian. 
The topological number $n$ of $U_{\rm F}$ in two dimensions is the Chern number of $\Gamma U_{\rm F}$. 
For a gapless mode of $H_{\rm F}$ at $\epsilon_{\rm F}=0$ $(\pi/\tau)$, 
the topological charge $\nu^{0 (\pi)}$ 
is the 1D winding number,
\begin{align}
\nu^\epsilon=\int_{s_{p\alpha}} \frac{\dd \bm{k}}{4\pi i} \cdot \tr \left[\Gamma (H_{\text{F}}(\bm{k})-\epsilon)^{-1} \nabla (H_{\text{F}}(\bm{k})-\epsilon) \right],
\quad (\epsilon=0,(\pi/\tau)),
\end{align}
where $s_{p\alpha}$ is a small circle surrounding the gapless point, and the branch cut of $H_{\text{F}}(\bm{k})$ is chosen at $\pi/\tau$ ($0$).
For simplicity, we set $\tau=1$ in the following.

\begin{figure}[htbp]
\centering
\includegraphics[width=140mm]{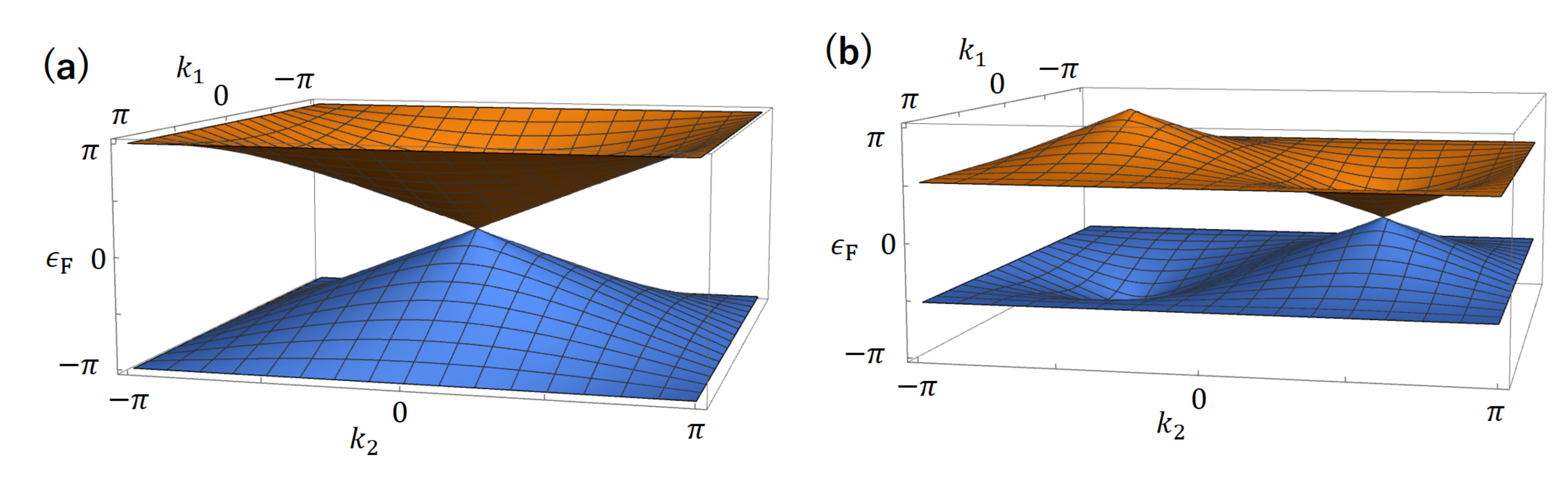} 
\caption{Floquet energy spectra of  $U_{\rm F}({\bm k})$ in Eq.~(\ref{eq:FAIII}) with
(a) $\theta=0$ and  (b) $\theta=3/4$.}
	\label{fig: FAIII}
\end{figure}

Let us consider the following 2D model in class AIII model, 
\begin{align}\label{eq:FAIII}
U_{\text{F}}(\bm{k}) =e^{i\theta \sigma_x/2}U_{y}^{-}(k_y/2) U_{x}^{-}(k_x) U_{y}^{+}(k_y/2) U_{y}^{-}(k_y/2) U_{x}^{+}(k_x) U_{y}^{+}(k_y/2)
e^{i\theta \sigma_x/2},
\quad \Gamma U_{\text{F}}^\dagger \Gamma^{-1}=U_{\text{F}},
\end{align}
where
$U_{j}^{ \pm}(k_j) =P_{j}^{ \pm} e^{\mp i k_{j}}+P_{j}^{\mp}$ with $P_{j}^{ \pm}=\left(\sigma_{0} \pm \sigma_{j}\right) / 2$ 
and $\Gamma=\sigma_3$.
When $\theta=0$, this model reduces to that considered in Ref.\cite{Higashikawa18}.
$\Gamma U_{\rm F}$ is written as
\begin{align}
\Gamma U_{\text{F}}
=d_x\sigma_x+ (d_y \cos \theta-d_z\sin \theta)\sigma_y+(d_z\cos\theta +d_y\sin\theta)\sigma_z
\end{align}
where
$d_x=-\cos^2 \left( k_x/2 \right)\sin k_y$, 
$d_y=-\sin k_x \cos^2 \left(k_y/2\right)$,
and
$ 
d_z=\cos k_x \cos^2 \left(k_y/2\right) -\sin^2 \left(k_y/2\right)
$. 
The Chern number of $\Gamma U_{\rm F}$ is evaluated as 1 for any $\theta$
as the vector $\bm{d}=(d_x,d_y,d_z)$ wraps the unit sphere once when ${\bm k}$ covers the Brillouin zone.

Figure \ref{fig: FAIII} shows the quasi-energy spectrum of this model with $\theta=3/4$.  
Near the Dirac point at $\epsilon_{\rm F}=0$, the Floquet Hamiltonian reads
\begin{align}
H_{\text{F}}({\bm k})\approx (k_x-\theta) \sigma_x + k_y \left(\cos^2 \frac{\theta}{2}\right) \sigma_y,
\end{align}
which gives $\nu^0=1$. 
On the other hand, 
near the Dirac point at $\epsilon_{\rm F}=\pi$, we have
\begin{align}
H_{\text{F}}({\bm k})\approx \pi + (k_x-\theta-\pi) \sigma_x - \left(\sin^2 \frac{\theta}{2} \right) k_y\sigma_y,
\end{align}
which gives $\nu^\pi=-1$. Thus, this model obeys Eq.~(\ref{eq:Thm2-2}) in Theorem 3'.


\end{document}